\documentclass[12pt,preprint]{aastex}

\usepackage{epsfig}
\def\Msun{M_\odot}
\def\kms{$\rm{km\,s^{-1}}$}
\def\arcsec{$^{\prime\prime}$}

\def\degree{$^{\rm o}$}
\def\CaII7291{Ca {\sc II}] $\lambda\lambda$ 7291,7323\ }

\def\OI6300{[O {\sc I}] $\lambda\lambda$ 6300,6364\ }

\def\Msun{M$_{\rm \odot}$}
\def\qisp{Q$_{ISP}$}
\def\uisp{U$_{ISP}$}
\def\fluxu{${\rm ergs\, cm^{-2} \AA^{-1} s^{-1}}$}
\def\dm15{$\Delta M_{15}$}
\def\degree{$^{\rm o}$}
\received{2002 June 21}
\begin{document}

\title {
Spectropolarimetry of SN 2001el in NGC 1448: Asphericity of a
Normal Type Ia Supernova\footnote{Based on observations collected at the
European Southern Observatory, Chile (ESO Progr. No. 68.D-0571(A).}
}

\author{Lifan Wang$^{1}$,
Dietrich Baade$^{2}$,
Peter H\"oflich$^{3}$, Alexei Khokhlov$^4$, J. Craig Wheeler$^3$,
 D. Kasen$^{1}$, Peter E. Nugent$^{1}$, 
 Claes Fransson$^5$, and Peter Lundqvist$^5$
}

\affil{$^1$Lawrence Berkeley National Laboratory 50-232\\
    1 Cyclotron Rd, CA 94720}

\affil{$^2$European Southern Observatory\\
    Karl-Schwarzschild-Strasse 2\\
     D-85748 Garching, Germany}
\affil{$^3$Department of Astronomy and McDonald Observatory\\
          The University of Texas at Austin\\
          Austin,~TX~78712}
\affil{$^4$Laboratory for Computational Physics and Fluid Dynamics, Naval
Research Laboratory, Washington, DC 20375; ajk@lcp.nrl.navy.mil}
\affil{$^8$ Stockholm Observatory, AlbaNova, Department of Astronomy\\
SE-106 91 Stockholm, Sweden}

\begin{abstract}
High-quality spectropolarimetry (range 417-860 nm; spectral resolution 1.27
nm and 0.265 nm/pixel) of the SN~Ia 2001el were obtained with the ESO Very
Large Telescope Melipal (+ FORS1) at 5 epochs.  The spectra a week before 
maximum and around maximum indicate photospheric expansion velocities of about 
10,000 km~s$^{-1}$.  Prior to optical maximum, the linear polarization of
the continuum  was $\approx 0.2 - 0.3 \%$ with a constant position angle, 
showing that SN~2001el has a well-defined axis of symmetry.  The polarization 
was nearly undetectable a week after optical maximum. 

The spectra are similar to those of the normally-bright SN~1994D with 
the exception  of a strong double-troughed absorption feature seen 
around 800 nm (FWHM about 22 nm).  The 800 nm feature is probably 
due to the Ca II IR triplet at very high velocities (20,000 - 26,000 \kms)
involving $\sim$ 0.004 \Msun\ of calcium and perhaps 0.1 \Msun\ total mass.
The 800 nm feature is distinct in velocity space from the photospheric
Ca II IR triplet and has a significantly higher degree of polarization 
($\approx 0.7 \%$), and different polarization angle than the continuum.  
Taken together, these aspects suggest that this high velocity calcium is a 
kinematically distinct feature with the matter distributed in a filament, 
torus, or array of ``blobs" almost edge-on to the line of sight.  This 
feature could thus be an important clue to the binary nature of SN~Ia, perhaps
associated with an accretion disk, or to the nature of the thermonuclear
burning, perhaps representing a stream of material ballistically ejected
from the site of the deflagration to detonation transition.

If modeled in terms of an oblate spheroid, the continuum polarization implies 
a minor to major axis ratio of around 0.9 if seen equator-on; 
this level of asymmetry would produce an absolute luminosity
dispersion of about 0.1 mag when viewed at different viewing angles.
If typical for SNe~Ia, this would create an RMS scatter of several
hundredths of a magnitude around the mean brightness-decline relation. 
We discuss the possible implications of this scatter for the
high precision measurements required to determine the cosmological 
equation of state.  

\end{abstract}

\keywords{stars: individual (SN 2001el) -- stars:
supernovae -- stars: spectroscopy -- stars: polarimetry}

\section{Introduction}
 
The last decade has witnessed an explosive growth of high-quality
data and models  for supernovae with spectacular results that provided new
perspectives for the use of SNe~Ia as cosmological yard sticks and for
constraining the physics of supernovae.
SNe~Ia have provided new estimates for the value of the Hubble constant
($H_0$) with 10 \%  uncertainty based on a purely empirical procedure  
\citep{Hamuy:1996,RiessPK:1996}, and on a comparison of detailed theoretical 
models with observations \citep{HofKho:1996,Nugent:1997}.
More recently, the routine successful detection of supernovae at
large redshifts, z \citep{Perlmutter:1997,Riess:1998},
provided results that are consistent with a low matter density in 
the Universe and, most intriguing of all, yielded hints for a 
positive cosmological constant, $\Omega_\Lambda $ $\approx 0.7$, 
and prompted the quest for the nature of the ``dark energy" or 
cosmological equation of state
\citep{PerlTurnWhite:1999}.
To pursue these issues with SNe~Ia, the required photometric accuracy has
to be better than 2 to 5 \% (Albrecht \& Weller 2000). 
This raises systematic effects as the main source of 
concern \citep{Hoeflich:1998}. In this context,
it is important to understand observationally whether or not SN Ia are
spherically symmetric since any deviations from sphericity could potentially
affect the accuracy of  distance estimates. 

Aside from the new focus on SN~Ia as cosmological tools, there are
long-standing problems associated with determining the progenitor
evolution and the physical processes involved in the explosion
by thermonuclear combustion. 
It is nearly universally assumed that SN Ia result from some form
of binary evolution, but no direct observational evidence for
this conjecture has ever been presented.  There is also general 
agreement that SNe~Ia result from some process involving the 
combustion of a degenerate white dwarf (WD) \citep{Hoyle:1960}. 
Within this general picture, three classes of models have been 
considered: (1) an explosion of a carbon/oxygen (CO) WD, 
with mass close to the Chandrasekhar limit, that accretes mass
through Roche-lobe overflow from an evolved companion star 
\citep{Whelan:1973}; (2) an explosion of a rotating configuration 
formed from the merging of two low-mass WDs, caused by the loss of 
angular momentum through gravitational radiation 
\citep{Webbink:1984,Iben:1984,Paczynski:1985}; and
(3) explosion of a low mass CO-WD triggered by the detonation of a helium
layer \citep{Weaver:1980,Wallace:1981,Nomoto:1982,WoosWeav:1986}. 
Only the first two models appear to be viable for observed SN~Ia 
\citep{HofKho:1996, Nugent:1997}. Although not favored for most 
SN~Ia, the merging of two WDs may contribute to the SNe~Ia 
population particularly, perhaps, to subluminous events \citep{Howell:99by}. 
All three scenarios might leave some trace in polarization data.  

There are at least three ways in which the fact that SN~Ia evolve
and explode in binary systems could impart a dominant axis
to the explosion that could be reflected in the polarization.  Perhaps
the most general is that the WD could be rotating. Rotation could
affect the shape of the WD, or it could affect the propagation
of the thermonuclear burning (see below) and hence the distribution
of either the elements produced or the density or both
\citep{Howell:99by}.  Another generic asymmetric feature
is an accretion disk.  Accretion disks resulting from mass 
transfer are not expected to be very massive and hence it is
difficult to see how the swept up disk matter could directly
affect the asymmetry. The exception might be if the disk comes from 
the disruption of a binary WD companion, in which case the remnant 
disk at the time of the explosion could be thick and dense.  
A third asymmetry generic to binary evolution models for SN~Ia is
a binary companion.  In two of the scenarios, the $M_{Ch}$ models and 
the helium detonators, the binary companion should still exist.  
The collision of the ejecta with this companion must induce some level 
of asymmetry in the ejecta \citep{Livne:1992,Marietta:2000}.

Other asymmetries, especially departures from a dominant
axis of symmetry could be intrinsic to the explosion process.
The departures from the dominant axis
could come from lumps of $^{56}$Ni that produce irregularities
in the excitation.  These, in turn, would be clues to the
nature of the thermonuclear burning that drives the explosion.
Within the class of $M_{Ch}$ models, it is believed that 
the explosion is triggered by compressional heating near the 
WD center, and that the burning front starts as a 
subsonic deflagration.  The time evolution of the burning front is
still an open question. The issue is whether the deflagration front burns
through the entire WD \citep{NTY:1984} or makes a transition into
a supersonic detonation mode, as suggested in the delayed detonation (DD)
model \citep{Khokhlov:1991,Yamaoka:1992}. 
DD models have been found to reproduce the optical and 
infrared light curves and spectra of ``typical" SNe~Ia reasonably
well \citep{Hoeflich:94D, HofKho:1996, Nugent:1997, Wheeler:IR,Lentz:2001}.  
The propagation of a detonation front is well 
understood \citep{Gamezo:1999,Sharpe:2001}, but the description of the 
deflagration front and the deflagration to detonation transition (DDT) 
pose problems.  

Significant progress has been made toward a better understanding of 
the physics of thermonuclear flames.  The front has been found to be 
Rayleigh-Taylor (RT) unstable, increasing the effective speed of the 
burning front (Nomoto et al. 1976) and also, perhaps, imposing density
irregularities (plumes) on the structure that could be reflected
in the polarimetry.  Starting from static WDs,
hydrodynamic calculations  of the deflagration fronts have been 
performed in 2-D \citep{Reinecke:1999} and 3-D
\citep{Livne:1993, Khokhlov:1995, Khokhlov:2002}. 
RT instabilities govern the morphology of the burning 
front and the effective burning speed is very sensitive to the 
energy release by the fuel, and therefore to the local C/O ratio
\citep{Khokhlov:2002}.  Therefore, the actual flame propagation
will depend on the detailed chemical structure of the progenitor.
Three-dimensional models \citep{Khokhlov:2002} suggest that
plumes of burned matter are frozen out in the expansion.  These
plumes, rich in $^{56}$Ni, could be the source of dispersion
around any dominant polarization axis.
On the other hand, optical and IR data on the sub-luminous,
but significantly polarized SN 1999by \citep{Howell:99by}, seem 
to be at odds with these models of the deflagration phase that 
predict significant mixing of the inner layers of the WD prior 
to detonation. Whether this conflict exists for normally bright
SN~Ia remains to be seen. The transition to detonation is 
thought to occur earlier in normally-bright events, giving 
less time for the RT structure to develop.  This might allow
clumps of $^{56}$Ni to develop that could drive the dispersion
in polarization vectors without causing intolerably large
distortions of the flux spectrum.
 
Pre-conditioning of the WD may affect the nature of both normal and
subluminous SNe~Ia. Such pre-conditioning may involve the main sequence mass
and metallicity of the progenitor WD
\citep{Dominguez:2001,Hoeflich:1998, Iwamoto:1999},
the accretion history \citep{Langer:2000}, large-scale velocity fields such
as turbulence prior to the runaway \citep{HofStein:2002}, or rotation
\citep{Howell:99by}.

A picture of the geometrical structure of different chemical constituents 
of different varieties of SN~Ia can thus be crucial for constraining the
physical models of the progenitor evolution and explosion.
During the first few weeks after the explosion, the photosphere 
recedes rapidly from the outermost stellar surface to deep inside the 
metal-rich central regions where the chemical composition is 
heavily affected by the explosion. Polarization is a powerful tool to probe
the structure of the rapidly expanding matter and to provide
detailed information about the geometrical structure of the ejecta that
is important in understanding supernova explosions and their subsequent
evolution to supernova remnants. Polarimetry can also be used to probe the
pre-explosion environment and the properties of the ISM along
the line of sight.

\citet{Wang:1996}, and \citet{WWH:1997} reported polarimetry data obtained at McDonald Observatory.
They studied supernova polarimetry published before
1996 and found that Type Ia supernovae are normally not polarized and on
average show much lower polarization than core-collapse supernovae
(Type II, Ib/c). The degree
of polarization of most SN Ia observed in \citet{Wang:1996} and \citet{WWH:1997} 
is unlikely
to be higher than 0.2-0.3\%\ at the observed epochs. The data also indicate
a higher degree of polarization for the bare-core or low mass events
such as Type IIb, and Ib/c supernovae, than for more massive
ejecta \citep{Wang:2001}. The degree of polarization is
a function of time after explosion.
For core-collapse events, the degree of polarization is higher past
optical maximum than before optical maximum
\citep{Wang:2001,Leonard:2000, Leonard:2001a,Leonard:2001b}.
These conclusions were based upon only a few events, and not all types of
supernovae were covered.

Spectropolarimetric observations of SN Ia and interpretation of the 
data are both very difficult as emphasized in \citet{WWH:1997}. 
Because of the large number of overlapping Fe II lines in an SN Ia 
atmosphere, polarization features of SN Ia are usually harder to
detect and interpret than those of SN II. For SN II, 
lines of H$\alpha$ and Na I/He I 5876 \AA\ normally
stand out and a geometrical picture can be built based on these
lines.  Further analysis of other chemical components can be constructed
following the structure deduced from the stronger lines.
\citet{WWH:1997} argue that despite the rather
noisy appearance of the spectropolarimetry of SN 1996X, polarization
of degree around 0.25\% may have been detected by comparing the
observed data with model calculations of SN Ia polarization spectra.
The only SN Ia published so far that showed a clear polarization 
signal of about 0.7\% is the sub-luminous event SN 1999by, which 
at optical maximum had a chemical structure very different than that of
normal  SN Ia \citet{Howell:99by}. It thus remains an extremely interesting
question to investigate the polarization of
spectroscopically normal SNe Ia, and other subclasses of SNe Ia.

In this paper, we present in \S 2 observations and data reduction of
SN 2001el.  We discuss the spectroscopic and spectropolarimetric 
evolution in \S 3 and \S 4, respectively.  The structure of the 
SN~2001el ejecta is studied in \S 5.  
In \S 6, we give a discussion and conclusions with emphasis on implications
for progenitor models of SN Ia and the effect of asymmetry on SN Ia luminosity.
Detailed discussions of the structure of the the ejecta with the focus
on the Ca II features are given in a seperate paper by \citet{Kasen:2003}.

\section{Observations and Data Reduction of SN 2001el in NGC 1448}

The bright SN Ia event SN 2001el provided a rare chance to study
the structure of a thermonuclear explosion. An image taken by the VLT
with FORS1 in imaging mode is shown in Figure 1. The SN was discovered
visually on Sept 17.064 UT at a magnitude about 14.5 \citep{Monard:2001}.
The host, NGC 1448, is a spiral galaxy nearly edge on with major axis
at position angle around 45\degree. Galactic reddening on the line of
sight to NGC 1448 is E(B-V) = 0.014 mag \citep{Schlegel:1998}.
The supernova is located at R.A. = 3h44m30s.60, Decl. =
-44o38'23".4 (equinox 2000.0), which is 14" west and 20" north of
the nucleus of NGC 1448.
The line of sight does not intersect the core or disk of NGC 1448 so that
extinction within the host galaxy should be small.
The SN was not detected down to 14.6 on Aug. 25.96 and was seen rising in
brightness the night after discovery \citep{Monard:2001}. 
The peak magnitude was V = 12.7 mag on about 1 Oct (N. Suntzeff
private communication).   It was classified
as a SN Ia well before maximum based on spectroscopic data taken using the
ESO-VLT Kueyen with the UVES, an echelle spectrograph, by 
\citet{Sollerman:2001} who also noticed that the characteristic 
Si II absorption at 612 nm appeared to be flat-bottomed.
The ESO UVES data show also interstellar absorption lines of Ca II H and K,
and lines of Na I D with equivalent width of 0.037 nm and 0.031 nm 
respectively, at the redshift (1162 kms$^{-1}$) of NGC 1448. The first set of
spectropolarimetry
data was obtained on Sept. 26 UT using the ESO-VLT Melipal with FORS1
\citep{Wang:2001b} for which we reported spectral features and
the positive detection of polarization at about 0.7\% -- comparable to what
is normally observed for Type II supernovae.

We subsequently decided to obtain spectropolarimetry of
SN 2001el in great detail for two purposes. First, we want to obtain data of
superb quality so that we can establish firmly the characteristics of
a definitely polarized SN Ia. This knowledge can then be used to assess the
degree to which we can trust other data sets -- both those observed from the
VLT and 
those obtained at smaller telescopes. Secondly, SN 2001el
for which the maximum luminosity was brighter than 13 mag provided a rare
opportunity 
to study in great detail the nature of SN Ia. We planned the observations
so that the evolution from before optical maximum to close to the nebular
phase was well covered. We also tried to guess the location of the
photosphere  while planning the observations
so that important epochs where the photosphere crossed a chemical layer
could
be probed. 

All of these 
observations were obtained
in service mode through our Target of Opportunity program at the VLT Melipal
with FORS1. FORS1 employs a TK2048EB4-1 2048$\times$2048
backside-illuminated
thinned CCD. The cooling of the CCD is performed by a standard
ESO bath cryostat. This system uses a FIERA controller for CCD
readout. The grism GRIS-300V (ESO number 10) was used for all
observations. The order separation filter GG 435+31 (4)  was used in some
observations
to study the Ca II IR triplet, but was also taken out to observe
spectral features in the blue end. The supernova was always observed
in four subsequent separate exposures with the position angle of the
Wollaston prism at 0\degree.0, 45\degree.0, 22\degree.5, and 67\degree.5.
The $\lambda/2$ retarder plates are of the ``superachromatic'' type, and the
position angles can be set with an accuracy of 0.1 degree. The chromatic
zero angles were set by the data provided in the VLT FORS1 user manual 
\citep{FORS1}.
The spatial scale is about 0.20\arcsec/pixel.
The spectral resolution for these observations is around 12.7
\AA\ as measured from the spectral calibration lamps -- with each pixel
corresponding to  about 2.6 \AA\ in wavelength scale. A slit of 1.0\arcsec\
wide is used for  all these observations.

A log of observations is presented in Table 1. In addition to the
observations
listed in  Table 1, we have also obtained images for bias correction,
wavelength calibration, and  flat fielding for each observing run. The
wavelength calibration and flat fielding  images are all obtained with the
waveplate rotated to the same position angles used  for the supernova
observations, i. e., 0\degree.0, 45\degree.0, 22\degree.5,
67\degree.5.
The flat field images of each run are combined to form the final flat field
correction image that is applied to each data set.
Standard data reduction procedures including bias correction, flat-fielding,
background subtraction, spectrum extraction, wavelength calibration, and
flux calibration were performed using the IRAF package \citep{Tody93}.
The polarized
spectra were reduced using our own software. The per-pixel photon
statistical errors of the Q and U vectors of the VLT data were typically
below 0.1\%. Because polarimetry involves the difference of the ordinary
and extra-ordinary light, the per-pixel errors are normally exaggerated and
are artificial because each resolution element contains more than
one pixel. It is always useful to rebin the data into bin sizes
comparable to the spectral resolution to reduce the artificial per-pixel
errors and error correlations. In this paper, all Stokes parameters were
rebinned to a 15\AA\ interval. The binning was done by calculating
the average polarization within a bin weighted by the number of detected
photons within each pixel \citep{WWH:1997}.  The 15 \AA\ binning was 
chosen so that it is slightly larger than spectral resolution of the 
data (12.7 \AA) to eliminate sampling errors of the resolution element. 
The rebinning is equivalent to calculating the degree of polarization 
of all photons integrated in each bin; when the bin size is larger 
than the resolution element, the errors of the Stokes Parameters in
each wavelength bin are uncorrelated.  Typical photon statistical errors 
in the Q and U vector after binning to the 15 \AA\ interval are 
around 0.02\% for all of these observations and are not a major 
source of error in the discussion of these data. During the 
commissioning of FORS1 in 1998/1999 the systematic instrumental 
polarization of FORS1 was found to be less than 0.1\% (Szeifert 1999, 
private communication). Errors during data reduction are expected to 
be small as well ($<0.02\%$) as the data are reduced in double precision.

\section{Spectroscopic Properties of SN 2001el}

Figure 2 shows the spectral evolution of SN~2001el. 
The date of $B$ maximum of SN~2001el was found to be Sept 31 by 
\citet{Krisciunas:2003}, from which we found that our spectra were
taken at -4, +1, +10, +19, and +41 days from $B$ maximum, respectively.
SN~2001el would be a spectroscopically normal supernova had the 
wavelengths beyond 750 nm not been observed in the pre-maximum phase. 
The match of spectral features to SN 1994D is extremely good except for 
the vicinity of the  Ca II IR triplet. 
The Ca II IR triplet is clearly detected in all our spectra of SN~2001el 
with a strength that grows in time at a velocity around -10,000
km~s$^{-1}$, close to the velocities given by other photospheric 
lines such as Si II 635.5 nm.  The feature that distinguishes
SN~2001el is a distinctively strong absorption at around 800 nm.
As argued below, we interpret this as a kinematically and geometrically 
separate filament or shell of calcium-rich material.

SN 1994D and other SN~Ia have shown weak features corresponding to 
the Ca II IR triplet at high velocities with a double-humped absorption
minimum \citep{Hatano:1999}.  Models of SN 1994D showed that the Ca II IR 
triplet can form a feature with a two-component absorption mimimum 
(H\"oflich 1995b; H\"oflich, Wheeler \& Thielemann 1998, their Figure 9), 
but that the double minimum structure is primarily 
dictated by the density and temperature through the 
ionization in such a way that the feature should recede in velocity as the 
ejecta expand.  Although the 800 nm feature fades with the approach
to maximum light, it always remains distinct from the photospheric Ca II IR 
triplet feature as the latter gains in strength.  The 800 nm feature
does not seem to evolve substantially in velocity space. In addition,  
the post-maximum photospheric component of the Ca II IR triplet in SN~2001el 
has a sharp blue edge at velocity around -12,500 \kms\ (see the last two 
spectra at +16 and +38 d in Figure 2). This implies that there is a sharp
density drop of Ca at that velocity. On the other hand, the pre-maximum
line profiles of the 800 nm feature show rather sharp edges that,
if identified with Ca II, correspond to a  velocity of about 
15,000 -- 20,000 \kms\ on the red side and about 26,000 \kms\ on the 
blue side (see the top spectrum in Figure 2).  The polarization data 
accentuates this delineation in velocity space as shown in the first 
panels corresponding to the data on 26 Sept in Figure 3.  
In any case, the red edge of the high velocity feature at the top
of Figure 2 does not overlap with the blue edge of the low velocity
Ca II feature at the bottom of Figure 2.  If both of these absorptions
are due to the Ca II IR triplet, this implies that the 
high-velocity filament or shell also has rather well-defined geometrical 
boundaries.  

The polarization data suggest that the lowest velocity matter in this 
high-velocity feature might be at about 17,000 \kms if the feature is Ca II.
The polarization shows that this feature also has 
a different geometrical orientation than the geometry that defines
the dominant axis of the photosphere.  Taken together, the velocity
separation, the large amplitude of the polarization, and the
different polarization angle all imply that this feature is a 
kinematically and geometrically separate high-velocity component
that is enriched in calcium.  

It is difficult to see whether something like this separate high velocity
feature exists in other SN~Ia.  The kinematic boundaries of the high-velocity 
calcium are impossible to discern in the data of SN 1994D, since the feature 
is so much weaker.  A search of published SN Ia spectra show that most 
of the SN~Ia with pre-maximum spectra covering the 800 nm area reveal weak 
spectral features similar to that of SN 1994D.  As for SN~1994D, however, it 
is difficult to discern whether this high velocity component is geometrically 
``detached" from the photospheric structure when the line is weak.  
In addition, 
the presence of Fe II absorption at about 800 nm can obscure the nature of 
this 
feature.  The current SN~Ia data base does seem to suggest that the very strong
feature, definitely displaced from the photospheric Ca II IR triplet,
is rather special to SN~2001el.
We address the possible physical origin of this feature in \S 5.

\section{Spectropolarimetry Properties of SN 2001el}

\citet{Wang:2001} outlined a method to decompose the observed polarimetry
into two components. On the Q - U plot, the two components correspond to
the polarized vectors projected onto the so-called dominant axis
and the axis perpendicular to the dominant axis. The dominant axis can
be defined from the aspherical
distribution of the data points on the Q - U plane.  The dominant 
axis is derived
by a linear fit to the data points weighted by the observational
errors in the Q - U plane. The spectropolarimetry projected to the
dominant axis represents global geometric  deviations from spherical 
symmetry whereas the 
vector perpendicular to the dominant axis represents deviations from the
dominant axis. The same method will be applied in this study as well.

We show in Figures 3 to 7 the observed data points in the Q - U plane. Each
point represent a data pair of the Q - U vector at a different wavelength. 
The wavelength of the data points in important intervals are encoded in color.
The data  show remarkable evolution during the five epochs of observation.
The polarization also shows spectral features that can be identified
with features in the flux spectra. This firmly
establishes that SN 2001el is intrinsically polarized,
at least at certain epochs after explosion.

SN 2001el exhibits some remarkable features that are unlike those of
previously observed SN II and the subluminous SN Ia 1999by
\citep{Wang:1996, WWH:1997, Wang:2001, Leonard:2001a, Leonard:2001b,
Howell:99by}. On the
Q - U plot, SN 1998S and SN 1999by showed well-defined linear features
\citep{Wang:2001,Howell:99by}. This is indicative of a relatively
well-defined
symmetry axes. The SN 2001el data, however, show large scatter around the
dominant axis. In particular, a sharp increase of the degree of polarization
is seen in the  Ca II IR triplet in the Sept. 26 and Oct. 1 data. 
Other strong
polarized spectral features are also observed during Sept. 26 and Oct. 1.
The polarized features becomes much weaker in the data taken on Oct. 18 and
afterwards.

\subsection{Interstellar Polarization}

To derive the intrinsic polarization due to the supernova atmosphere, we
first
need to deduce the component due to interstellar dust. A simple approach
is to assume that the resonance-scattered photons are unpolarized 
\citep{Trammell:1993}. This method attempts to 
distinguish continuum and scattered photons
and use that separation to derive interstellar extinction
\citep{Jeffrey:1991,Trammell:1993,Hoeflich:1996w,
Wang:1996, Tran:1997, Leonard:2001a}. This technique implicitly assumes a
unique intrinsic polarimetry axis and the derived interstellar polarization
should fall on the dominant axis. The interstellar polarization derived from
this method,  however, is normally inconsistent with the initial
assumptions.
In the case of SN 1993J,  it is also clear from the data in the Q - U plane
that
there are significant deviations from the dominant axis \citep{Wang:2001}.
Another method is outlined in
\citet{Wang:2001} that relaxes the constraints on the polarization
properties
of emission lines  and assumes only that at each specific epoch, the
polarization produced by the supernova ejecta has a single polarization
position angle  independent of wavelength. This method would set the
interstellar polarization at one of the two ends of the dominant axis. This
method again, is perhaps a useful method only for those data with a
well-defined dominant axis.

Neither of the above two methods are applicable for the SN Ia data to be
discussed
in this paper. The time evolution of the polarization data in fact show
that the degree of polarization decreases significantly after
maximum light.
The data taken on 2001 Nov. 9, in particular, show flat Q and U spectra with
little 
evidence of deviations across most of the strong lines. At this phase,
the supernova has entered the nebular phase, and there are substantial
reasons to believe that the ejecta have become optically thin to electron
scattering. Intrinsic polarization due to the supernova ejecta should be
small
at this phase. Panels (b) and (d) of Figure 7 show that there are perhaps
still
detectable polarized features at the P-Cygni absorption dip of the Ca II IR
triplet, but the degree is significantly smaller than
for earlier epochs. In
wavelength regions excluding the Ca II IR triplet feature, the fluctuations
of
the polarization are about 0.1\% -- consistent with earlier observations on
other post-maximum SN Ia. If we assume that at this epoch
in the wavelength range 430 --
500 nm the supernova is intrinsically unpolarized, we find the
interstellar polarization to be at $Q\ =\ 0.01 \pm 0.05$, and $U\ = \ 0.60
\pm
0.05$. We will adopt this as the best estimate of the interstellar
polarization.

The Galactic extinction is E(B-V) = 0.014 \citep{Schlegel:1998}, 
which would produce
a maximum degree of polarization of 0.13\% \citep{Serkowski:1975} --
too small to account for
the 0.6\%\ interstellar polarization. According to \citet{Serkowski:1975},
the observed interstellar polarization implies that the total extinction to
the supernova is larger than E(B-V)\ = \ 0.067, with the extinction by the
host galaxy being larger than 0.053.


The wavelength dependence of the interstellar polarization gives reasonable
estimates of the properties of the dust particles on the line of sight
to the supernova. A fit to the 
empirical formula 
\citep{Serkowski:1975, Wilking:1982} with the Sept. 9 data binned to 
60 nm is shown in Figure 8. The fit shows that the wavelength at which
the polarization reaches maximum is $\lambda_{max}\ = \ 514\ \pm\ 10 $ nm. 
For comparison, the Galactic value of $\lambda_{max}$ is around 550 nm.
The fit implies that the dust that is responsible for
the interstellar polarization is similar to that of the Galaxy.
Adopting the relation $R_v\ = (5.6\pm0.3)10^{-4}\lambda_{max}$ 
\citep{Serkowski:1975, Larson:1996, Whittet:2001}, we found that
$R_v = 2.88\pm0.15$. 

The above analysis yields a polarization position
angle of 45\degree\ for the interstellar polarization.
This implies that the magnetic field is parallel to the
disk of the host galaxy (cf. Figure 1).

\subsection{Decomposition of the Polarization}

With the above choice of interstellar polarization, we can now decompose the
observed polarization into the two components as described in the 
introduction to \S 4. The new coordinates are given by rotating 
the original coordinate systems counter-clockwise so that the Q-axis
overlaps the dominant axis in the new coordinate system. The dominant
axis is determined to be at position angle $\alpha$ = -26.57\degree\ 
through a linear fit to the data points on  the Q - U plot of Sept. 26. 
We have rotated the Q - U coordinate system by 26.57\degree\ clockwise 
so that the dominant axis points
toward the center of the  data cluster on the Q - U plot. The components
parallel and  perpendicular to the dominant axis are given by

$$
P_d \ = \   (Q-Q_{ISP}) \cos \alpha\ + \ (U-U_{ISP}) \sin \alpha, \eqno(1)$$
and
$$P_o \ = - (Q-Q_{ISP}) \sin \alpha\ + \ (U-U_{ISP}) \cos \alpha, \eqno(2)
$$
where $P_d$ and $P_o$ are the polarization components parallel to
the dominant axis and orthogonal to that axis, respectively, and
\qisp\ and \uisp\ are the Stokes parameters of the interstellar
polarization.

To understand the physical meaning of the new coordinate system, it is useful 
to start from the simple spheroid models described in \citet{Hoeflich:1996w}
and \citet{WWH:1997}. In those models, the polarization is produced by electron
scattering through an aspherical atmosphere, and the spectropolarimetric
lines are formed  because of line scattering. Such a geometry would produce
a single line on the  Q - U plot, with the wavelength region that suffered the
most line scattering being most de-polarized and hence the closest to the
origin. Such an ideal system apparently does not  apply to SN 2001el where,
as can be clearly seen, the Q - U vectors show  significant scatter around
the dominant axis. The spheroid model can be modified to account for the scatter. 
For example, chemical clumps and filaments are
reasonable sources for deviations from the single axis geometry.

Under the modified spheroid scenario, the dominant axis most likely measures
the global departure of the photosphere from spherical symmetry. In the case
of an axially symmetric spheroid, $P_o$ measures the deviations and
fluctuations from  axial symmetry.

$P_d$ and $P_o$ are plotted in Figure 9 where it can be seen that the $P_o$
components are flat for these observations except for the spectral features
due to a few strong lines. Most of the continuum polarization is recorded in
$P_d$. The wavelength at which the degree of polarization reaches maximum
was around 740 nm on Sept. 26, moved to about  670 nm on Oct. 1, and  
to 550 nm on Oct.  9. The $P_d$ distribution became practically flat 
on Oct. 18  and Nov. 9.  The fact  that the Oct. 18 and Nov. 9 
polarization spectra are flat again supports our choice of interstellar 
polarization.  The polarized flux of the dominant component $P_d$ 
peaked at 650 nm, 610 nm, and 550 nm on Sept. 26, Oct. 1, and Oct 9, 
respectively, and became flat on Oct. 18 and Nov. 9. For the 
spheroid model of \citet{Hoeflich:1996w} and \citet{WWH:1997}, 
the blueward evolution of the peak of the polarized flux 
indicates that the last scattering region  was moving toward zones 
with increasing electron temperature.  Polarized spectral  
features are unambiguously detected around Si II 635.5 nm and the Ca
II IR triplet, especially in the first two peaks.

\subsection{Polarization of the Si II 635.5 Line}

The evolution of the Si II 635.5 nm line is shown in Figure 10.
On Sept. 26 and Oct. 1, before optical maximum, the
Si II 635.5 line is clearly polarized. On the
plot for Sept. 26, $P_d$ shows a peak in the velocity range from
-10,000 \kms\ to  0 \kms, with a sharp drop at the -10,000 \kms\
edge. 
The $P_o$ component is practically flat, but with a broad weak dip from
around -20,000
\kms\ to +10,000 \kms. On the Q - U plot, the data points of the Si II line
form 
an ellipse elongated along the dominant axis. Similar polarization behavior
is seen in Figure 10b. The degree of polarization decreased to less than
0.2\% on data taken on Oct. 9 and afterwards.  Note the displacement of the
matter
moving at +20,000 \kms\ from that at -20,000 \kms\ and the suggestion of a
``loop" in the Q-U plane for the matter from 0 to -10,000 \kms\ 
in Figure 10a.  In Figure 10b the highest velocity components have 
merged in the Q - U plane, but the suggestion of a
``loop" for the 0 to -10,000 \kms\ component remains.

The feature centered in the range -25,000 \kms\ to -20,000 \kms\
is especially interesting.  It shows peak-to-valley polarization 
variations of 0.3\% and 0.2\% on Sept. 26 and Oct.  1, respectively. 
This feature could be reasonably attributed to a neighboring line; but it
is also consistent with being the silicon counter-part of the high-velocity
feature observed most clearly in the Ca II IR triplet.

\subsection{Polarization of the Ca II IR Triplet}

The strongest polarization is observed for the high velocity Ca II IR
triplet, as shown in Figure 11. The polarization around 800 nm
reached 0.7\%\ on Sept 29 and Oct 1 while the high
velocity component in the flux spectrum was still strong.  
The polarization decreased in strength rapidly with the redward 
evolution of the high velocity component in the total flux spectrum.
The polarization data suggest that the lowest velocity matter in this
high-velocity feature might be at about 17,000 \kms\ if the feature is Ca II.
On Oct. 9, the Ca II line  close to the photosphere (at -12,500 \kms) becomes
spectroscopically stronger than the  high-velocity component. 
It is worth noting that no peculiar evolution of the polarization 
feature was observed to be associated with the emergence
of the  lower velocity Ca II feature.

In the Q - U plane the orientation, especially of the high velocity feature,
is at an angle of about 60\degree\  to the principal axis in the 
Sept. 29 and Oct. 1 data.  This means that the high
velocity calcium represents a very distinct geometrical as well as velocity
component in the explosion, with the difference in the orientation
being about 30\degree\ in the object.  Taken together, the velocity
separation, the large amplitude of the polarization, and the
different polarization angle all imply that this feature is a
kinematically and geometrically separate high-velocity component
that is enriched in calcium.

\section{The Structure of the SN 2001el Ejecta}

The polarization evolution allows us to extract some useful information 
about the ejecta structure.  For simplicity, we will assume that close 
to the photosphere the ejecta can be modeled as a prolate or oblate spheroid. 
We define the asphericity of the spheroid as the ratio of the major to
minor axes ratio minus 1, expressed as a percentage.  This spheroid 
reproduces the dominant axis of asymmetry.  The photosphere itself 
may be irregularly structured, so the spheroid is only an approximation.
The relation between this geometric asymmetry and the angular 
distribution of the luminosity for an electron-scattering atmosphere 
is taken from \citet{Hoeflich:1991} and \citet{WWH:1997}.

\subsection{Continuum Polarization}

For the maximum degree of polarization to be 0.7\%\ on Sept 26 and Oct 1,
an asphericity of ~25\%\ would be required for an ellipse observed along the
equator \citep{Hoeflich:1991, Howell:99by}. 
The 0.7\%\ was, however, observed only for the 800 nm feature associated
with the high-velocity Ca II IR triplet.  If we think that the
high-velocity component does not represent the global geometry of the
photosphere, the maximum degree of polarization would be 0.2 -- 0.3\%\ 
from the mean polarization spectrum on the dominant axis 
$P_d$ shown in Fig. 8 (a), (b), and (c).  This is considerably
smaller, but would still require asymmetries around 10 -- 15 percent at the
photosphere.  The asphericity would be larger if the ellipsoid 
were viewed in a direction other than along the equator.  
This asymmetry could be produced by a distortion in the density
distribution or in the excitation by $\gamma$-ray deposition
\citep{WWH:1997, Howell:99by, Hoeflich:2001}. This amount of asymmetry
could lead to luminosity-viewing angle  dependence with dispersion 
in brightness on the order of 0.1 magnitude even for otherwise
identical events. We return to this possibility in \S 7.

Small scale inhomogeneities in the chemistry, density, or excitation
must be present to account for the fact that the data points on 
the Q -- U plot do not fall on a single line, but show significant 
scatter around the dominant axis.  Clumps of radioactive $^{56}$Ni 
at the photosphere could produce these small scale, random components 
as irregularities in the excitation.  It would be interesting to try to
model the number of clumps necessary to generate this granularity, but
since the polarization represents a vector addition of the individual
components, this is not a trivial excercise. We leave it for the future.
  
Further insight into this structure that deviates from the
dominant axis may be obtained from the variation of the polarization
angle with time and wavelength.  With the adopted interstellar polarization, 
the principle component, $P_d$, changes sign with wavelength from 
blue to red.  In the pre-maximum spectra (see the first two panels of 
Figure 9).  A flip of the polarization vector can occur if the location 
of an energy source changes with respect to the photosphere. 
As discussed for SN~1993J \citep{Tran:1997, Hoeflich:93Jpol}, 
if the $^{56}$Ni source is well below the photosphere, the flux is radial,
but the flux acquires a tangential component if the energy source is close 
to the photosphere.  This ``skin effect" can cause a flip in the 
polarization angle.  In 2001el, the flip of $P_d$ with wavelength may 
be the result of such a ``skin effect" in combination with the 
frequency-dependent opacity. In the blue, the photosphere is formed at  
much larger radii (by a factor of 2) compared to the red. 
If an excitation blob is buried in the blue but revealed in the red, 
the polarized flux will be determined by the radial and tangential 
flux, respectively, with an associated difference in the orientation
of the polarization.

In SN Ia, this flip of the sign of the polarization along the
dominant axis might be due to a single large $^{56}$Ni clump 
revealed at the photosphere in the red \citet{Hoeflich:93Jpol}.
Such a single dominant blob would have to fall off the dominant
geometric axis to produce the sign flip.  Alternatively, there
could be a distribution of a number of small blobs that are on
average concentrated along the preferred axis.  In this case the 
sign flip could occur when a number of these clumps are exposed,
first in the red, by the receding photosphere.  If this were the
case, both the dominant polarization axis and the dispersion
around it might be attributed to a common origin.

About 1 to 2 weeks after maximum light, the photosphere has receded well
into the $^{56}$Ni rich region and, consequently, the effects of anisotropic
excitation and ionization will vanish as the $\gamma$-ray deposition
becomes smooth \citep{Hoeflich99byIR}.  The polarization is then 
expected to become small, consistent with the observations (see
the last panel in Figure 9).

\subsection{Line Polarization}

The most dramatic aspect of the polarization of SN 2001el is the
feature at 800nm.  This behavior of the  Ca II IR triplet may 
provide a key to understanding the departure from
axial symmetry.  Among the possibilities to explain the strong 
high-velocity Ca II IR triplet in absorption at 22,000 \kms\ are 
a spherical shell, dense clumps, some of which fall along the line 
of sight, or a torus that intersects the line of sight to the 
photosphere of the supernova.  As noted in \S 3, the sharp edge to
the photospheric component of the Ca II IR triplet, implies that
the density of Ca drops off beyond the photosphere before the
high-velocity component is encountered.  The high-velocity component
is thus a distinct geometrical component, not simply a monotonic
continuation of the calcium to high velocity in the ejecta.

As remarked in \S 3, the high-velocity Ca II absorption line 
seems to be especially strong in SN 2001el.  
The strength in SN 2001el could be due to physical differences in this
one event, but it could also be that this high-velocity feature 
is ubiquitous, that its presence is transient, and that its 
amplitude is sensitive to viewing angle.  
If the high-velocity Ca II is distributed aspherically in all SN~Ia, 
then SN~2001el may have just presented us with a lucky line of sight.  
There are relatively few cases in which observations have been
made before maximum light, but in that situation, some evidence for
a high-velocity feature is often seen, but evidence for a feature as 
strong and as disconnected in velocity space from the photospheric 
component as that in SN~2001el is, so far, unique to this event.

The existence and strength of the pre-maximum Ca II high-velocity 
component provides constraints on its physical nature.  
Whether the high-velocity Ca II component in SN 2001el is a single 
structure like a filament, a torus, dense clumps or some other geometry, 
it could partially obscure the photosphere and make the photosphere 
appear asymmetric in the absorption lines.
This would add another component to the polarization caused by an
intrinsically aspherical photosphere.  If this structure were an 
axially-symmetric torus or a prominent filament, 
the axis of symmetry would have to be different from the symmetry 
axis of the photosphere to account for the orientation of the 
high-velocity calcium feature in the Q - U plane.  If this component 
were caused by a calcium-rich clump, then the clump must be displaced 
from the principal axis.  If the feature is caused by a distribution
of calcium-rich clumps, then this distribution must deviate 
significantly from the dominant axis of the photosphere.  

If the interpretation of the high-velocity component as a torus 
were correct, then, statistically, one expects a correlation between 
the strength of the high-velocity absorption component and the 
size of its polarization.   This would still be true for an 
asymmetric distribution of clumps, for instance if a surrounding
torus were itself clumpy, but would not be true if the clumps
responsible for obscuring the photosphere were themselves distributed
isotropically around the explosion so there were no net effect
of orientation angle.

The Si II 635.5 nm line falls predominantly along the $P_d$ axis in 
the Sept. 29 data (Fig. 9a) when the high velocity Ca shows a distinct
orientation along the $P_o$ axis.  The fact that in the Q -- U plot, 
the Si II line also forms a ``loop" on Sept. 26 and Oct. 1 shows that 
the Si II 635.5 nm line also cannot be explained only by an 
axially-symmetric spheroid.  The Si ``loop" might show some
preference to an orientation along the $P_o$ axis in the Oct. 1 data 
(Figure 10b).  This suggests that there might be a high velocity 
Si counter-part to the high velocity Ca, but the evidence is not
conclusive; nevertheless, it is possible that the Si II line 
shares the same geometry as the Ca II line.  If true, this would
be an important constraint on the physical nature of the
high-velocity feature.

\section{Physical Models for Asymmetries}

How, then, might one account for a principle polarization
axis of SN 2001el, the dispersion about that axis and,
especially, the detached, high-velocity feature seen 
so strongly in SN 2001el?  
Detailed modeling with 3-D radiative transfer is required to
fully understand the polarization data \citep{Hoeflich:1996w, WWH:1997,
Howell:99by}. This analysis will be presented elsewhere. Here
we will discuss some of the physical possibilities.  

As outlined in the Introduction, there are a variety of ways
that systematic asymmetries could be imposed on the explosion of
a SN~Ia:  rotation of the exploding star, an accretion disk, 
the presence of a binary companion, or plumes of combustion
products.  Some of these could define the dominant axis, 
others could account for varying orientation axes and still
others for the dispersion observed around the dominant axis.
Some affects of rotation have been sketched in \citet{Howell:2001}
and we will not repeat them here.  
\citep{Marietta:2000} show that the impact of the supernova ejecta 
with the secondary star, assumed to fill its Roche lobe, creates a hole
in the ejecta with an angular size of $\sim 30$\degree\ in the 
high-velocity ejecta and with an angular size $\sim40$\degree\ 
in the low-velocity ejecta, or 7\%-12\% of the ejecta's surface. 
This effect might be able to induce the small photospheric
polarization along the dominant axis that we report here, and
it might not be observable in all SN~Ia due to orientation effects.
We will return to the possible effects of an accretion disk below. 

The greatest challenge in the current observations is to account for
the high-velocity shell of calcium-rich material. The
low-velocity Ca II seems to share the velocity, the degree of 
polarization, and the polarization angle with the photosphere.  
By contrast, the high-velocity matter defined by the 800 nm feature
differs from the photosphere in velocity (by definition), in the degree
of polarization, in the polarization angle, in the optical thickness, 
and in the filling factor.  If the Ca II identification is correct,
the high-velocity matter is physically and geometrically detached 
from the lower-velocity material on the basis of the sharp edges of 
the absorption lines.  The depth of high velocity feature requires
that there must be two zones of Ca II. These zones differ by so much in 
geometry, dynamics and column density that it is difficult to accommodate 
them in an exploding star of only one component.  The impressive 
homogeneity of SN~Ia does not leave much room for major individual 
peculiarities.  This is an additional constraint every model of 
SN~2001el needs to satisfy.  We will sketch here and critique several 
possibilities.

\subsection{The Mass of High-Velocity Ejecta}

In order to make a high-velocity shell of calcium-rich material
modestly optically thick so that it can obscure a patch of the
photosphere and induce a significant polarization feature, there
must be a substantial mass of high-velocity calcium.  A simple
static atmosphere argument suggests the following.  Taking an
effective opacity of order 1 cm$^2$ gm$^{-1}$ and shell radius
and thickness of order $10^{15}$ cm, the mass of calcium must
be of order $5\times10^{-3}~f\kappa^{-1}R_{15}^2$ \Msun\ to 
give an optical depth of order unity, where $f$ is the filling
factor of the high-velocity shell.  

A more rigorous approach is to consider the Sobolev optical
depth of the line:
\begin{equation}  
\tau = \chi c \rho \frac{dr}{dv},
\end{equation}
where dr/dv = t in a homologously expanding atmosphere
and $\chi$ is the integrated line opacity given by
\begin{equation}
\chi = \frac{\pi e^2}{m_e c}\frac{f_{ul} n_l}{\nu_{ul} \rho}
\left(1 - \frac{g_l n_u}{g_u n_l}\right),
\end{equation} 
where $f_{ul}$ and $\nu_{ul}$ are the oscillator strength 
and frequency for the line that represents a transition form
the lower (l) to upper (u) levels.  Neglecting the second term in
parenthesis involving the statistical weights and writing the 
population in the lower level as $n_l = n_{Ca}~exp(-E_l/kt)$, 
where $n_{Ca}$ is the total number of calcium atoms, 
the Sobolev optical depth can be written as:
\begin{equation}
\tau = \frac{\pi e^2}{m_e c} f_{ul}~n_{Ca}~\lambda_{ul}~t~e^{-E_l/kt},
\end{equation}
where $E_l$ is the energy of the lower level.  With $f_{ul} \sim 3$,
$E_l = 60,533$ cm$^{-1}$ = $1.2\times10^{-11}$ erg, and a temperature
of about 5000 K, the Sobolev optical depth can be written as:
\begin{equation}
\tau = 2.6\times10^{15} \rho_{Ca}~t_6,
\end{equation}
where $t_6$ is the time in units of $10^6$ s.
The mass of calcium in a spherical shell with filling factor, f, 
can thus be written:
\begin{equation}
M_{Ca} = 0.002~M_\odot \frac{f \tau R_{15}^2 \Delta R_{15}}{t_6}.
\end{equation}
For a putative Ca line at 20,000 \kms\ with a width of $\sim 5000$ \kms,
the mass of calcium would thus be $M_{Ca} = 0.004~ f \tau t_6^2$ M$_\odot$. 
For optical depth about unity and filling factor of unity, we again
need of order 0.01 \Msun\ of calcium to form the high-velocity
feature.

In regions of the white
dwarf where Ca is formed by incomplete Si burning, the 
layers are typically 60 percent Si, 30 percent S and 3 percent Ca
\citep{Hoeflich99byIR}.  If this composition is typical of the 
high-velocity shell, then the total mass of this matter might be of 
order 0.1 $f \tau t_6^2$ \Msun.  As remarked earlier, 
if the paucity of this strong, well-separated feature 
in many SN~Ia is an indication of 
the covering factor, then $f$ could be substantially less than unity.  
Given the relative strengths of observable lines, it is plausible that 
the Ca would give the strongest observed line, with Si weak, and 
S difficult to observe, as may be the case for SN~2001el.

\subsection{The Geometry of High-Velocity Ejecta}

A related perspective is to consider the shape and depth of
the high-velocity feature.  The feature is approximately flat
bottomed (ignoring the double minimum in the 26 Sept spectrum
of Figure 2). This suggests that the line is optically thick 
over a substantial velocity range.  The line should thus
be saturated if the feature obscured the whole photosphere.
From the depth of the line (about 30 percent of the adjacent
continuum flux), we deduce that the geometry can be blocking
no more than about 30\degree\ of the photosphere.  Note
that if the high-velocity component extended to higher latitudes,
there should be a low-velocity component along the line of 
sight.  The rather sharp edges of the high-velocity feature
and its lack of overlap with the photospheric component of
the Ca II IR triplet again suggest that the portion of
this high-velocity feature with finite optical depth does
not blanket the photosphere.  Some sort of torus geometry
could satisfy this constraint.  To account for what may
be common weaker high-velocity features in SN~Ia, some of
the high-velocity calcium may have to be distributed more
isotropically, but with modest optical depth.  If this
component is present in SN~2001el, then it must be in a
manner that low-velocity wings on the high-velocity feature
are weak. Whether a self-consistent model can be developed 
will require more rigorous consideration with multi-dimensional
transfer.

\subsection{The Origin of High-Velocity Ejecta}

One possibility is that this high-velocity matter is 
produced in the explosion in the region of partial Si burning.
In this case, a stream of matter must be ejected from relatively
deep within the explosion as, perhaps, a filament of high-velocity
matter.  This might happen through some sort of ``pinch" effect.
A particularly interesting possibility is that this high velocity
feature represents an artifact of the process of deflagration to
detonation transition.  The physics of this process is not yet
well understood, but one possibility being explored is that the
transition occurs between a rising plume of burned material and a 
sinking plume of fuel.  Ignition of a detonation at the sheet-like
interface between these structures might, in some circumstances, dynamically
eject a ballistic blob of matter that were subject to partial
Si burning.   The hint that some silicon may share the high velocity
kinematics of the calcium would be consistent with the composition
expected in this picture.  If this were the case, the high-velocity feature
would be an important clue to the burning dynamics. 

Another intriguing possibility is that the 
high-velocity matter is a clue to the binary nature of 
the progenitor.  In particular, it is clear that any matter
in an accretion disk should be swept up into a high-velocity
layer, possibly maintaining a toroidal structure.  The question
then arises as to how this matter could be calcium-rich.  

The most commonly envisaged scenario would have a hydrogen-rich
disk.  The collision of the most rapidly moving outer parts of
the WD ejecta, exceeding 10,000 \kms, would heat any such disk to 
high temperatures.  If the post-shock conditions were optically
thick and dominated by radiation pressure, the temperature
would be of order $1.2\times10^8$ K $\rho^{1/4}v_9^{1/2}$ where
$v_9$ is the shock velocity in units of $10^9$ cm s$^{-1}$.  If
the post-shock matter can radiate so that the shock energy goes
into purely the thermal motion of particles, then the temperature
could be as high as $4\times10^9 v_9^2$ K.  The density
is expected to be low, less than 1 gm cm$^{-3}$, so while the 
post-shock temperature could be sufficiently high that burning
might occur, the burning timescale could be too long compared to 
the dynamical timescale, which will be of order a second at the inner
edge of the disk at the boundary with the WD.  If it occurs, 
the burning in this hydrogen-rich situation would correspond
to the fast rp-process \citep{Wallace:1981}.  Burning would break 
out of the CNO cycle at this high temperature and proceed to higher
atomic weight matter.  The resulting composition may not
be calcium-rich, but dominated by more neutron rich species.  

Another interesting possibility is
that disk is composed of helium, the product of transfer
from a helium star.  Such a disk might be denser, depending
on the the temperature and transfer rate \citep{Cannizzo:1984},
and the composition would allow burning along the 
standard $\alpha$-chain with the possible production of
Ca, along with Si and S, through incomplete Si burning.
While it is clearly premature to reach any conclusions
regarding the nature of this high-velocity Ca feature,
the prospect that it represents the shock burning of 
a helium-rich accretion disk would give both direct
indications that the explosion occurred in a binary
system and that a helium, not hydrogen, companion were involved.

\section{Discussion and Conclusions}

 Understanding the 3-D nature of SNe~Ia is an important key to the
underlying physics, and polarimetry is the tool to probe 
multi-dimensional effects.

 The current observations provide strong evidence that the normal Type Ia SN
2001el is polarized before optical maximum. The polarization cannot 
be explained by a purely axially symmetric configuration; rather 
several components are required: 1) an axially symmetric component with 
polarization of about 0.2 to 0.3 \%  with a well-defined
axis.  This component indicates that the overall asphericity
of the photosphere is around 10 percent and perhaps more, depending
on viewing angle.  This global asymmetry may be produced by either
asymmetric excitation and ionization or it may reflect asymmetry
in the overall density distribution.
2) A component to account for the spread in the position angle 
with wavelength around the principal axis in the Q - U plane  
that indicates additional structure with more random orientation. This
component could be caused by clumps in the chemical structure.  
If the clumps are related to the $^{56}$Ni distribution, they could
be related to irregularities in the excitation expected from the RT
instabilities during the deflagration phase of the explosion
\citep{Khokhlov:2002}. 3) The high degree of polarization in the 
high-velocity component of the Ca II IR triplet.
This component, which is at an angle of about 60\degree\ 
to the principal axis 
in the Q - U plane, can be understood qualitatively as an obscuration of 
a portion of the photosphere by a filament, clumps,  or
perhaps a torus-like structure expanding at high velocity. 
The sharp blue edge of the photospheric component of the Ca II
IR triplet and the relatively sharp red edge of the high-velocity
component indicates the origin of the high-velocity component 
in a separate structure well separated from the photosphere.  
The fact that the line is not saturated suggests that the 
filling factor is substantially less than unity.

Excitation due to the non-uniform distribution of $^{56}$Ni provides 
a natural explanation for observations of the photospheric polarization, 
and hence clues to the burning process, but other explanations are 
possible.  The merger scenario, would, for instance, predict a 
global asymmetry due to the large expected rotation of the merged
object.  The merger scenario is not likely for SN~1994D, which
showed evidence for explosive carbon burning in the outer layers 
(i.e. IR Mg II lines), but no evidence for unburned carbon 
\citep{Bowers:1997, Wheeler:IR} as would be expected in 
a merger scenario.  Since SN~2001el is in many ways similar to
SN~1994D, a similar argument might be applied, but the IR
observations were not made.  It may be that detailed analysis
and models can address this issue from the polarization and
other data at hand for SN~2001el.

The polarization data provide interesting perspectives on the physics of SN
Ia phenomena, but there are many open questions. What is the origin of 
the high velocity Ca II in SN~2002el? 
Could the Ca II shell be the relic of an 
accretion disk? If so, what is the composition of the accretion disk 
and could it be burned into Ca? We will focus on these issues 
in separate studies incorporating detailed theoretical modeling.
The model-dependent interpretations of the high-velocity CaII feature 
could be put on a much firmer basis if future observations confirm 
a) a correlation between the strength and the degree of polarization of 
the Ca II IR triplet, b) the $\sim$ 30\degree\ difference in
the polarization angle of the continuum and of the high-velocity matter,
and c) a correlation between the polarization of the continuum and 
the polarization of the high-velocity Ca II, presuming their respective 
dominant axes are defined by the equator and the rotation or magnetic axis.

Finally, we return to the implications of asymmetry for the use of
SNe Ia for cosmology.  A 10 percent asymmetry of the photosphere 
would not cause systematic difficulties for using SN Ia as distance 
indicators at the current level of accuracy of about 20 percent.
This level of asymmetry would, however, cause a directional dependence 
of the luminosity of order 0.1$^{\rm m}$ \citep{Hoeflich:1991}, 
and a corresponding, but smaller, dispersion in the brightness-decline 
relation of SNe~Ia.   This dispersion depends on the viewing-angle 
dependence of the luminosity variation and, thus, the nature of 
the asymmetry.  The angle dependence of the luminosity of a single
SN~Ia will not, in general, vary as the line of sight to
the equator, that is, as cos$\theta$.  Thus, even in a large
sample, this effect is not expected to average out.  Whether
the remaining bias is slightly positive or slightly negative
depends on the specifics of the asymmetric luminosity distribution.  
Whether this bias can be eliminated to the few percent level needed 
to determine the cosmological equation of state by applying 
the \dm15\ correction or by employing the CMAGIC method 
\citep{Wang:2003} remains to be seen. We note that if such
effects are present, then SN~Ia are even more homogeneous than
they seem from current dispersions in peak brightness.  In
principle, if the angle-dependent luminosity could be determined
and removed quadratically from the data, the dispersion
could be reduced from current values.    

More observations are required to understand the polarization behavior 
of SN Ia and its effect on large samples of supernovae.  Once it is
known from polarization statistics the phases in which the polarization
is small, more weight can be assigned to the ``low-risk" phases
of the light curve.  Even if there remains a bias due to the
angle-dependence of the emitted flux, this may have no affect in
a large sample as long as the bias is independent of redshift.  The
relative changes in peak magnitude with redshift that are required
to determine the equation of state of the ``dark energy" would
not be affected by such a constant bias.  Even then a redshift
dependent bias might be small if the only effect were that
of an ellipsoidal variation of rather small amplitude for which
the luminosity variation with angle were small and gentle.
The pernicious question is whether there is an effect on the
polarization and hence luminosity distribution that was a rather
strong function of observer orientation and that also depended on
redshift. An example would be polarization induced by the collision
of the supernova ejecta with a companion star that might only
be seen from a limited range of observer viewing angles coupled
with SN~Ia being dominated nearby by SN~Ia in systems with
non-degenerate companions and at large redshifts by mergers
that left no companion, or vice versa.   Further study of the
polarization of SN~Ia should elucidate this issue.

\noindent{\bf Acknowledgements:} The authors are grateful to the European
Southern Observatory for the generous allocation of observing time. 
They especially thank the staff of the Paranal Observatory for 
their constant and untiring support of this project in service 
mode.  This work was supported in part by NASA Grant NAG5-7937 to PAH
and by NSF Grant AST 0098644 to JCW.

\clearpage
\begin{deluxetable}{r|rrrrrr}
\tablecolumns{7}
\tablecaption{Log of Observations \label{tbl-1} }

\tablehead{Star  & Date           & Exposure & Wavelength (nm)          & Airmass
& Grism     & Filter }

\startdata

\tableline
\tableline
SN 2001el & 2001-09-26.284 & 4$\times$60s & 418.1 -- 863.5 & 1.11 &
GRIS-300V & GG 435 
\\
SN 2001el & 2001-09-26.297 & 4$\times$480s & 418.1 -- 863.5 & 1.10 &
GRIS-300V & GG 435 
\\
\tableline
SN 2001el & 2001-10-01.205 & 4$\times$500s & 418.1 -- 863.5 & 1.30 &
GRIS-300V & GG 435 
\\
\tableline
SN 2001el & 2001-10-09.240 & 60.0s & 333.0 -- 863.4 & 1.12 & GRIS-300V
& --   
\\
SN 2001el & 2001-10-09.243 & 4$\times$240.0s & 333.0 -- 863.4 & 1.11 &
GRIS-300V & -- 
\\   
SN 2001el & 2001-10-09.261 & 4$\times$400.0s & 418.1 -- 8635 & 1.10 &
GRIS-300V & GG 435 
\\
\tableline
SN 2001el & 2001-10-18.161 & 4$\times$500s & 418.1 -- 8635 & 1.29 &
GRIS-300V & GG 435 
\\
SN 2001el & 2001-10-18.191 & 4$\times$500s & 333.0 -- 8634 & 1.18 &
GRIS-300V & --  
\\
\tableline
SN 2001el & 2001-11-09.101 & 4$\times$1200s & 333.0 -- 863.4 & 1.29 &
GRIS-300V & --  
\\

\enddata

\end{deluxetable}

\clearpage


\figcaption{The field of SN 2001el and the host galaxy NGC 1448.
The SN is marked by the arrow.
}


\figcaption{Spectral evolution of SN 2001el. The spectra are similar to those
of normal SN Ia except for the strong absorption line at around 800 nm.
Spectra of SN 1994D (dotted lines) are also shown for comparison. 
The phase of the SN 1994D data are day -4, +2, +11, and +20 from top 
to bottom. The absorption line
at 800 nm is seen in pre-maximum spectra of SN 1994D as well, but is
much weaker than that of SN 2001el. This line could be due to
a high velocity component of the Ca II IR triplet. 
}


\figcaption{
Spectropolarimetry of SN 2001el on 2001 Sept. 26, 7 days before maximum. The
Stokes parameters are rebinned into 15 \AA\ bins. An interstellar
polarization component is subtracted from the observed Stokes Parameters so 
that the data points represent intrinsic polarization due to the supernova. 
The assumed interstellar polarization is shown as a solid dot in the 
Q - U plot (panel a - upper left).  Without subtraction of the interstellar 
component, the origin of the coordinates would be centered at this solid dot. 
The straight line illustrates the dominant axis shifted to the origin of 
the Q - U plot.  The Q (panel b - upper right) and U (panel d - lower right) 
spectra show conspicuously polarized  spectral features. The degree of 
polarization is shown as the thin line in panel c (lower left) with the flux 
spectrum (panel c, lower left, thick line) overplotted to show the correlations 
of the degree of polarization and the spectral features.  The wavelength color 
code is presented at the bottom of panel a.
}


\figcaption{
Same as Figure 3, but for data taken on Oct. 1, 2 days before maximum.
}


\figcaption{
Same as Figure 3, but for data taken on Oct 9, 7 days after maximum.
}


\figcaption{
Same as Figure 3, but for data taken on Oct 18, 16 days after maximum.
}


\figcaption{
Same as Figure 3,  but for data taken on Nov 11, 38 days after maximum,
with the data rebinned to 50 \AA. }


\figcaption{Flux, polarized flux ($F\cdot P_d$ and $F\cdot P_o$),
$P_d$ and $P_o$ of SN 2001el on Sept. 26 (a), Oct. 1 (b), Oct. 9 (c),
Oct. 18 (d), and Nov. 11 (e). The flux is in units of 1 $\times\ 10^{-14}$
\fluxu. The polarized fluxes are in units of 1 $\times\ 10^{-16}$ \fluxu.
}


\figcaption{Fit to the interstellar dust polarization using the
empirical Serkowski law $p_\lambda/p_{max}\ = \ 
exp[-K ln^2(\lambda_{max}/\lambda)]$ \citep{Wilking:1982, Serkowski:1975}.
The data are from Sept. 9 observations.
}


\figcaption{ The evolution of polarization signatures of Si II 635.5 nm. The Si
II 635.5 line shows conspicuous polarization on Sept. 26 (a) and Oct 1 (b). The
polarization signature weakened considerably for data taken on Oct. 9(c) and is
undetected for data taken Oct. 18 (d) and Nov. 11 (e)}


\figcaption{ The evolution of the polarization signatures of the Ca II IR
triplet. The velocities are calculated with respect to rest wavelength 860 nm.  
The Ca II IR triplet is polarized on Sept. 26 (a) and Oct 1 (b). The 
strongest polarization is observed for the component with velocities 
around -25,000 \kms\ and -20,000 \kms.  The polarization signature 
weakened considerably for data taken on Oct.  9(c), Oct. 18 (d), and Nov. 11 (e)}

\newpage
\plotone{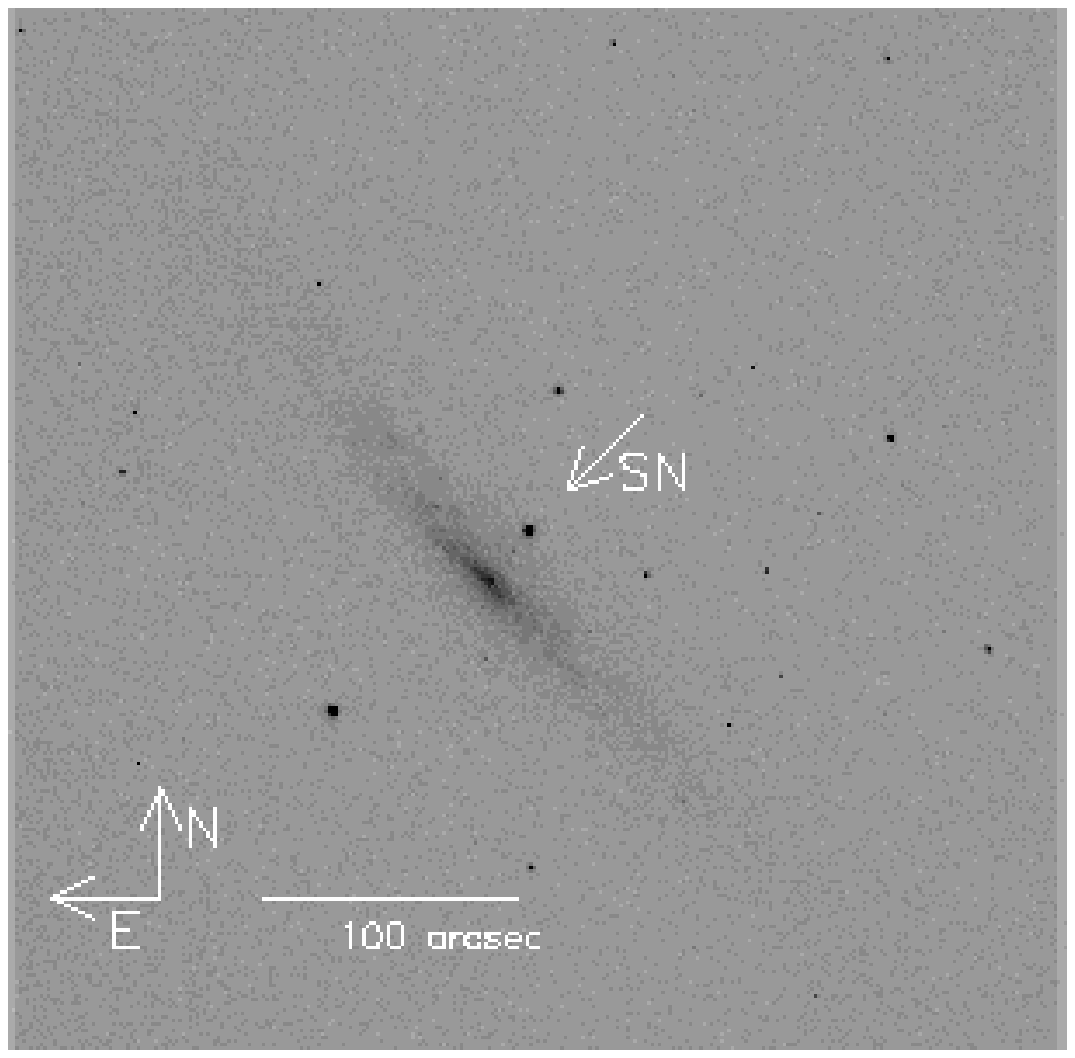}
\newpage
\plotone{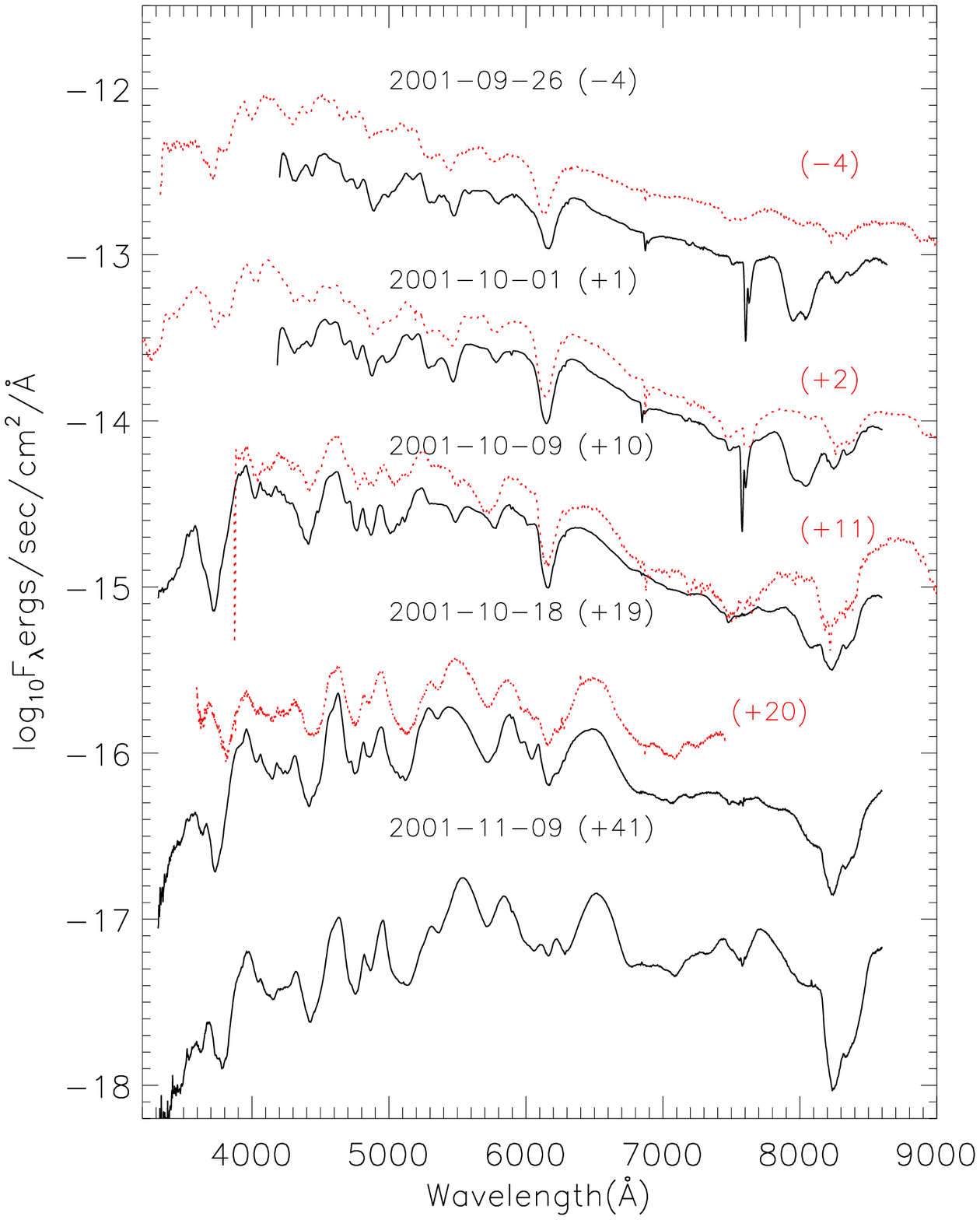}
\newpage
\plotone{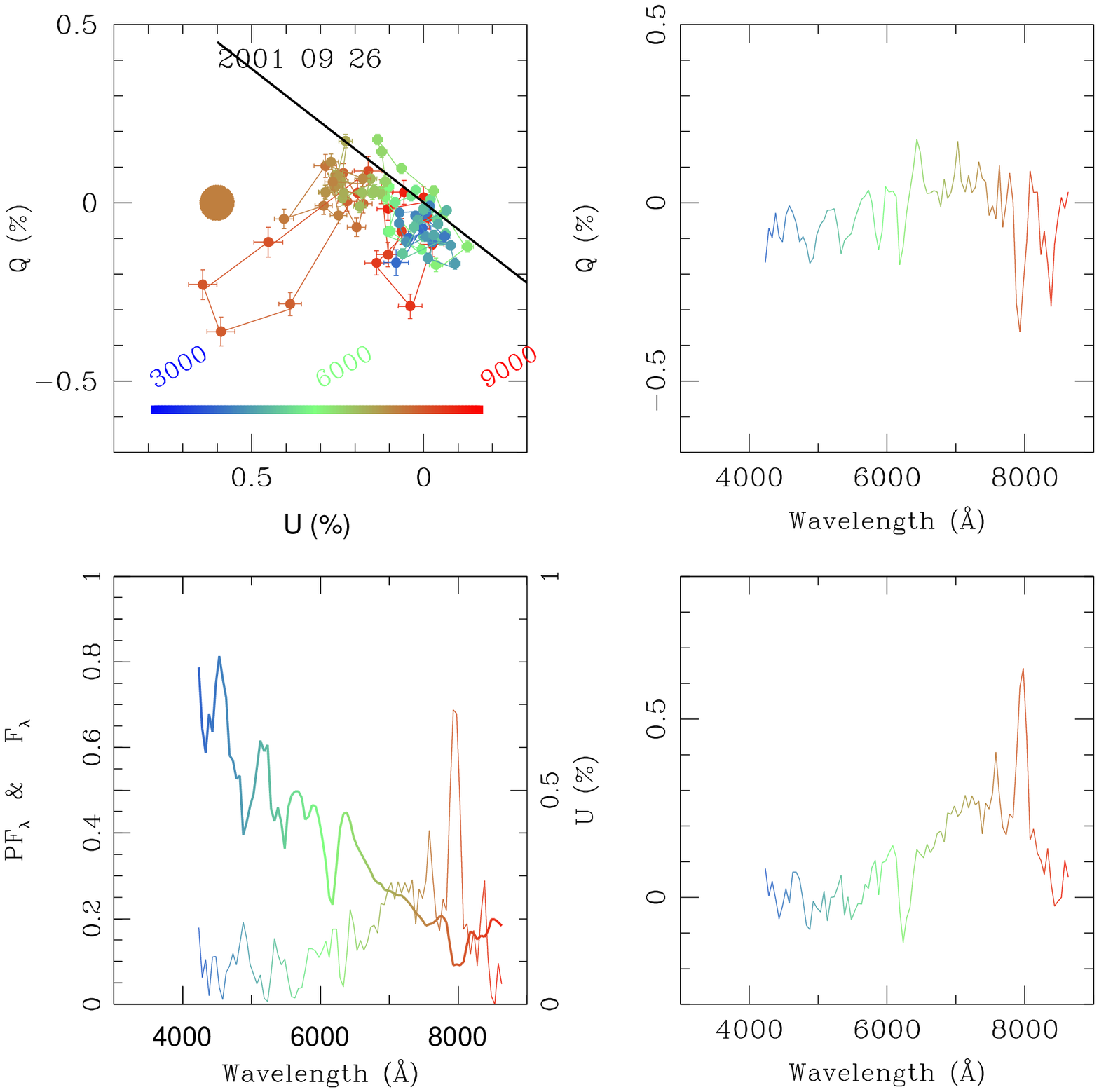}
\newpage
\plotone{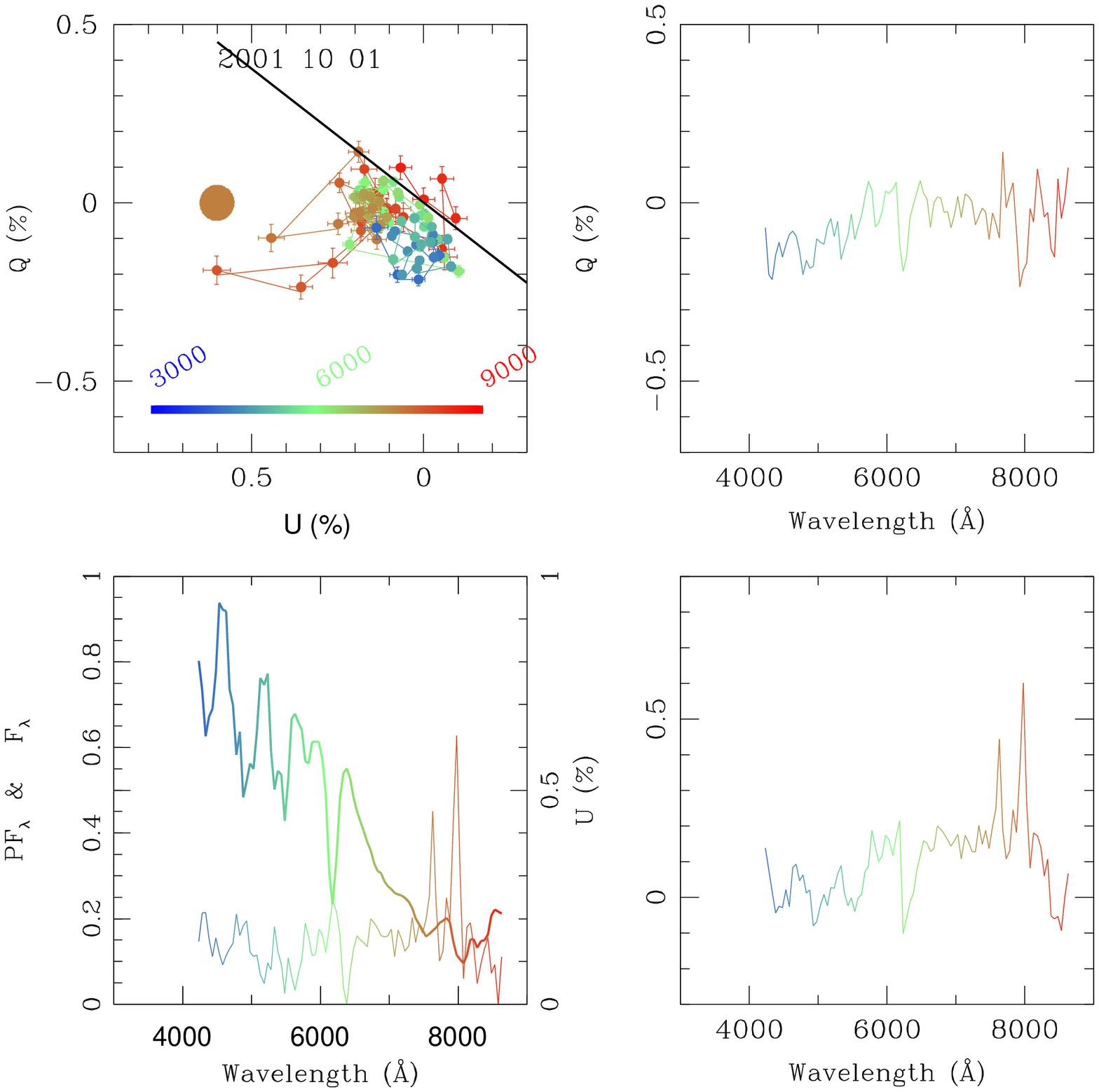}
\newpage
\plotone{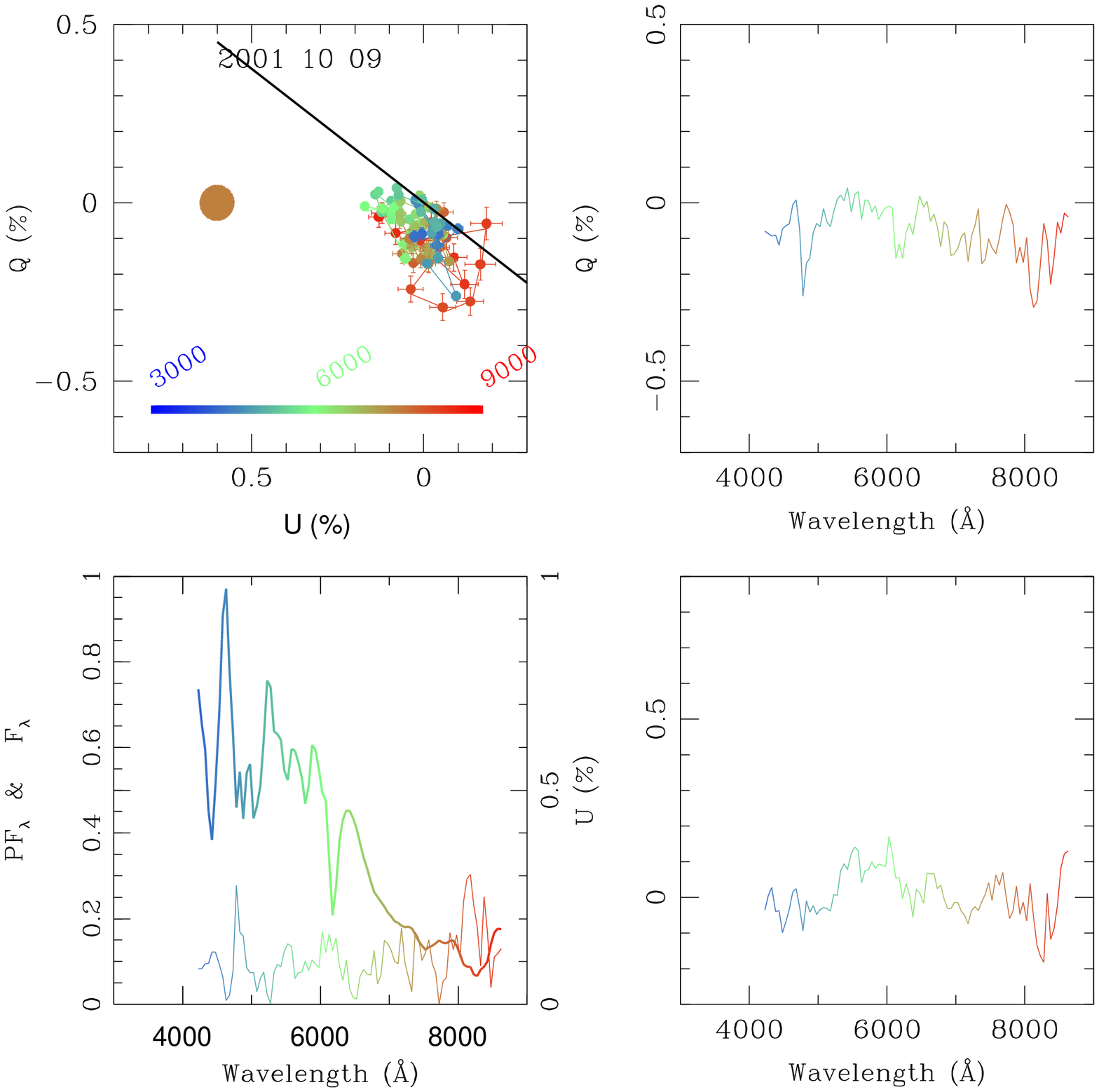}
\newpage
\plotone{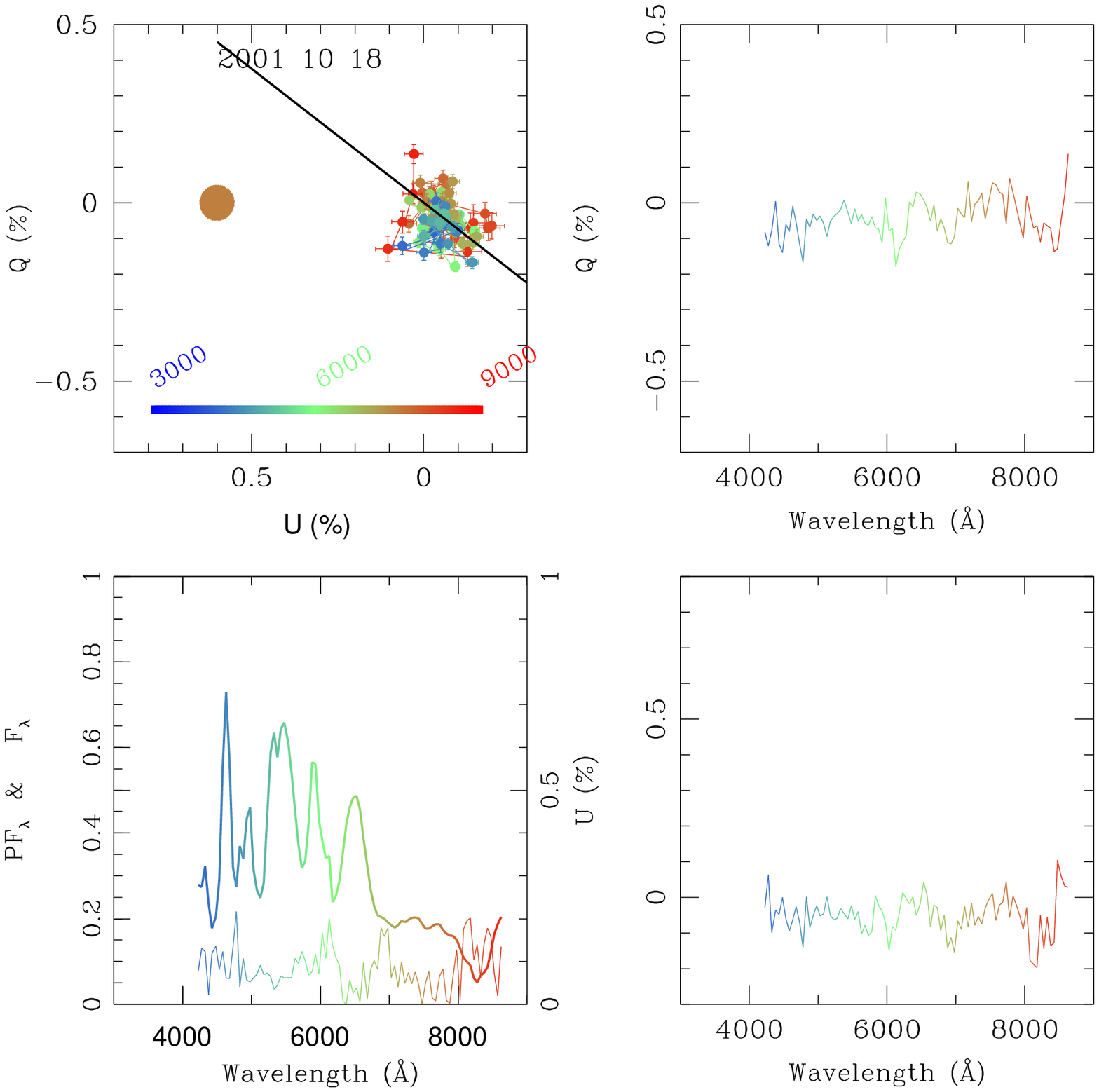}
\newpage
\plotone{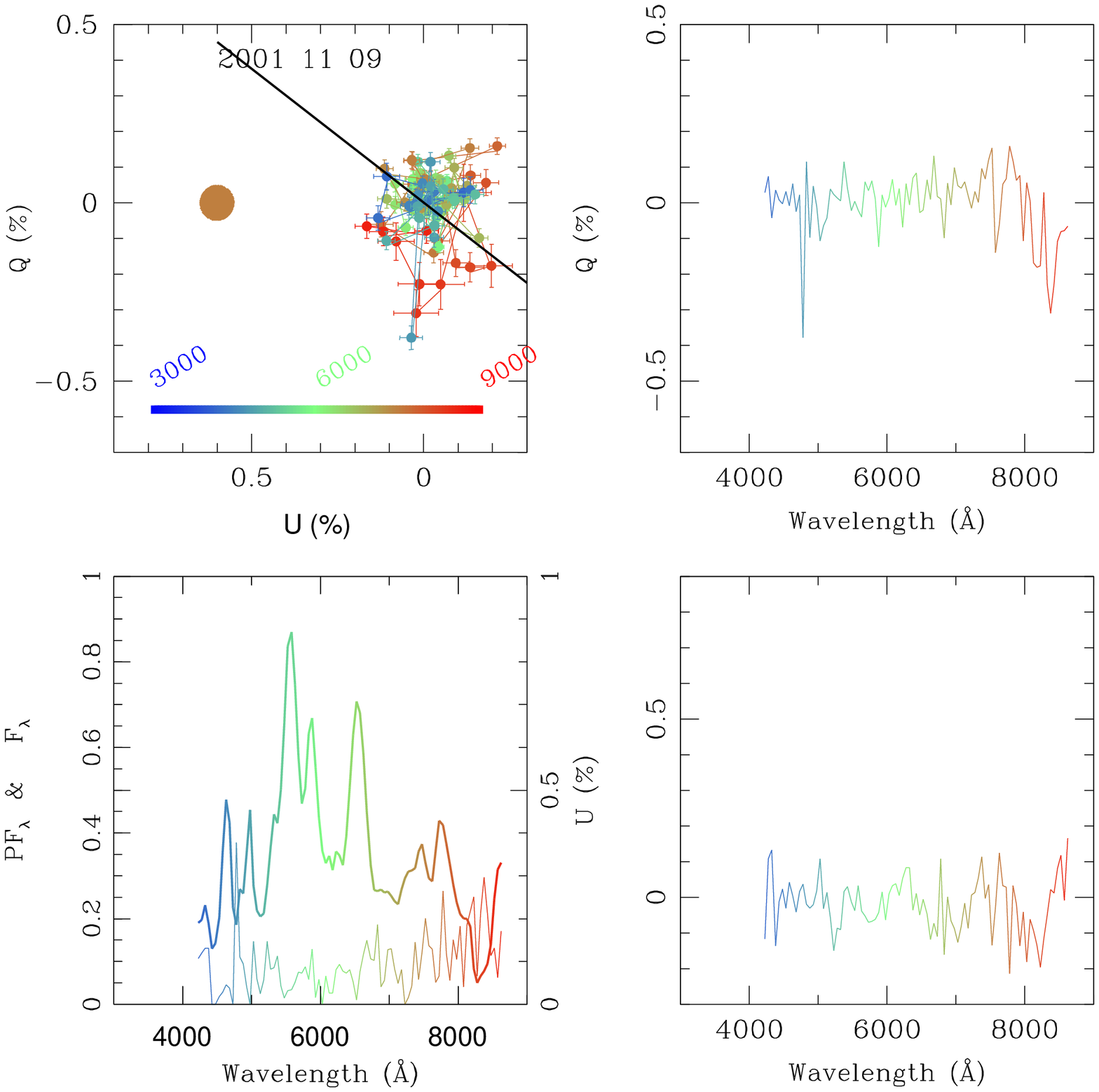}
\newpage
\plotone{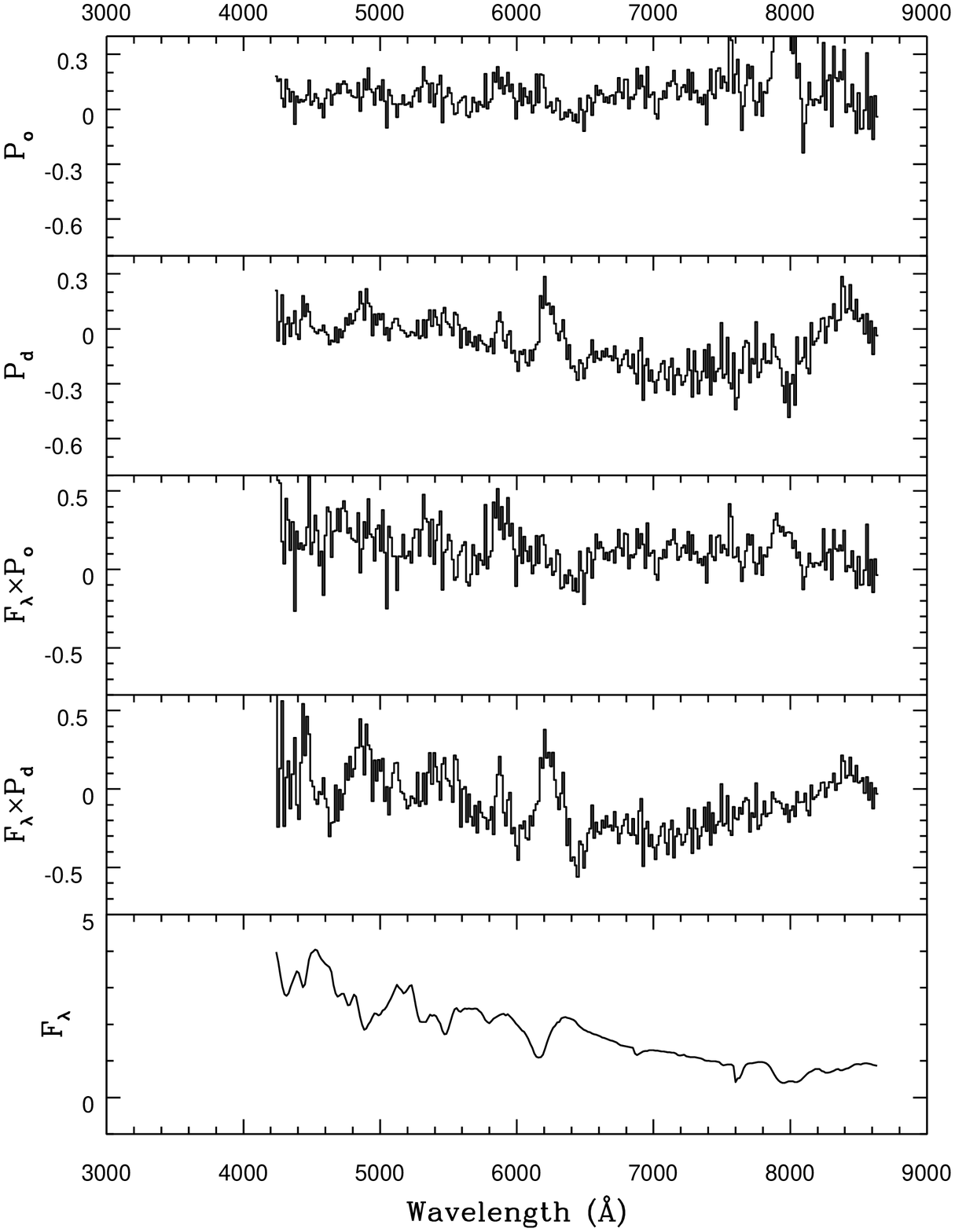}
\newpage
\plotone{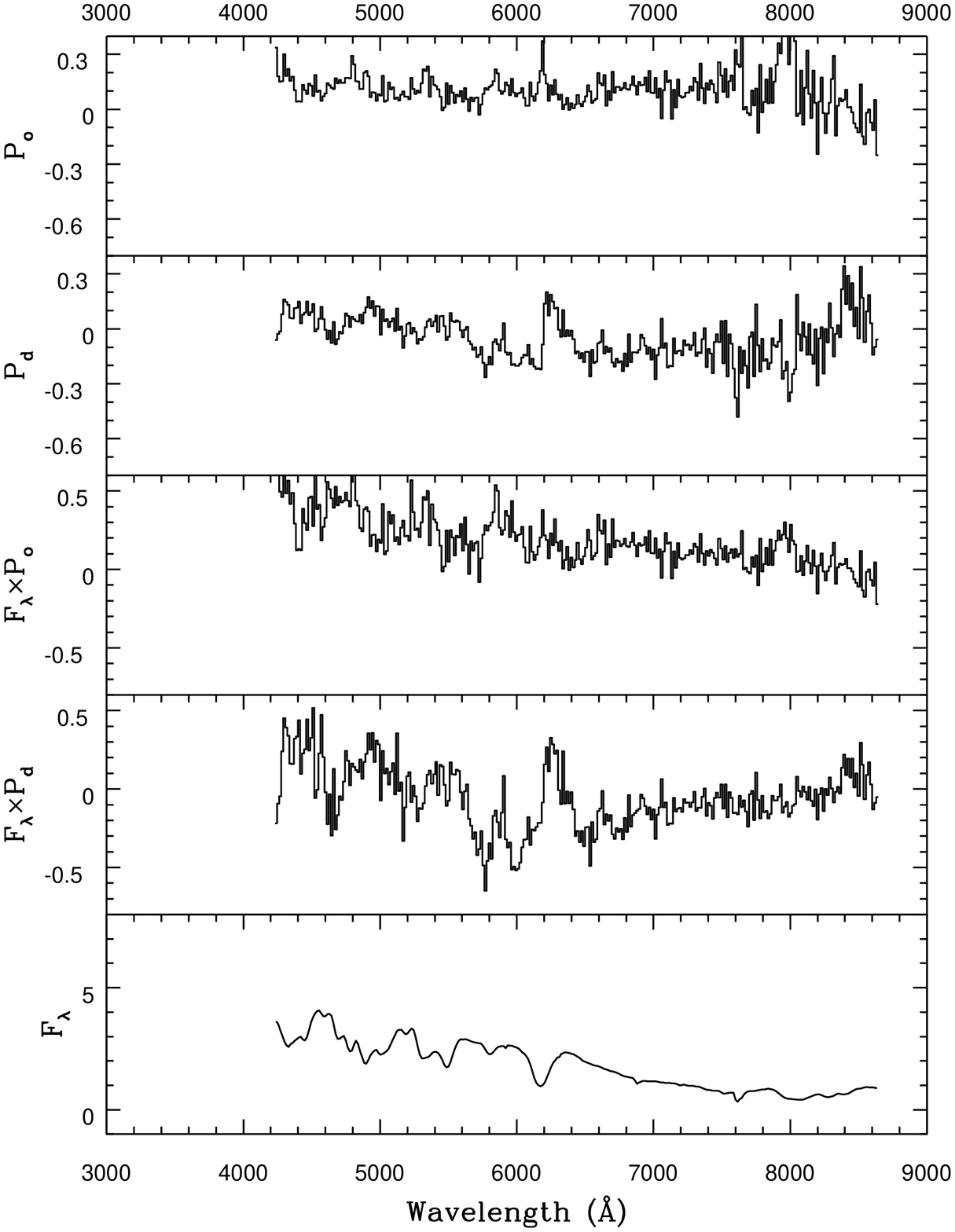}
\newpage
\plotone{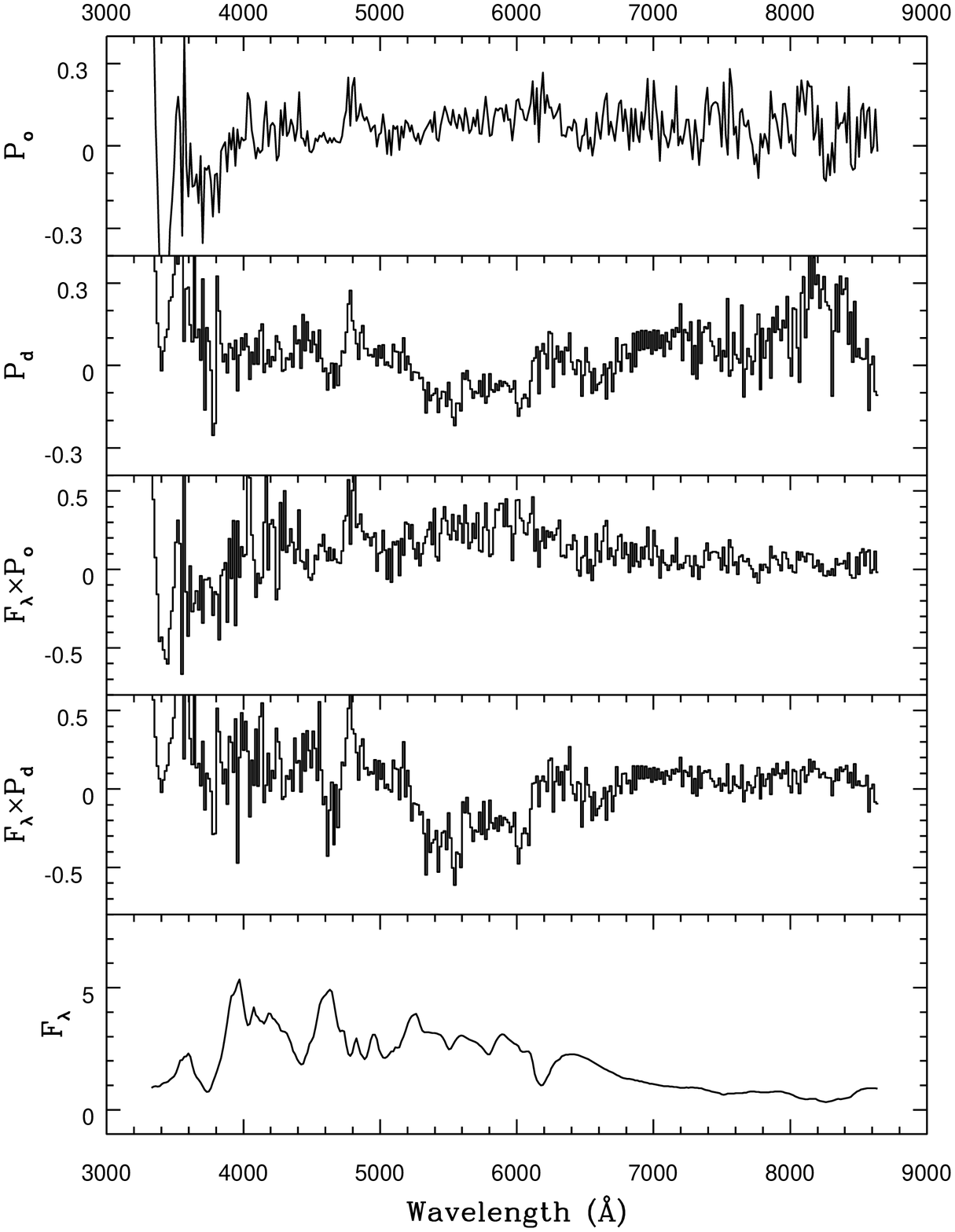}
\newpage
\plotone{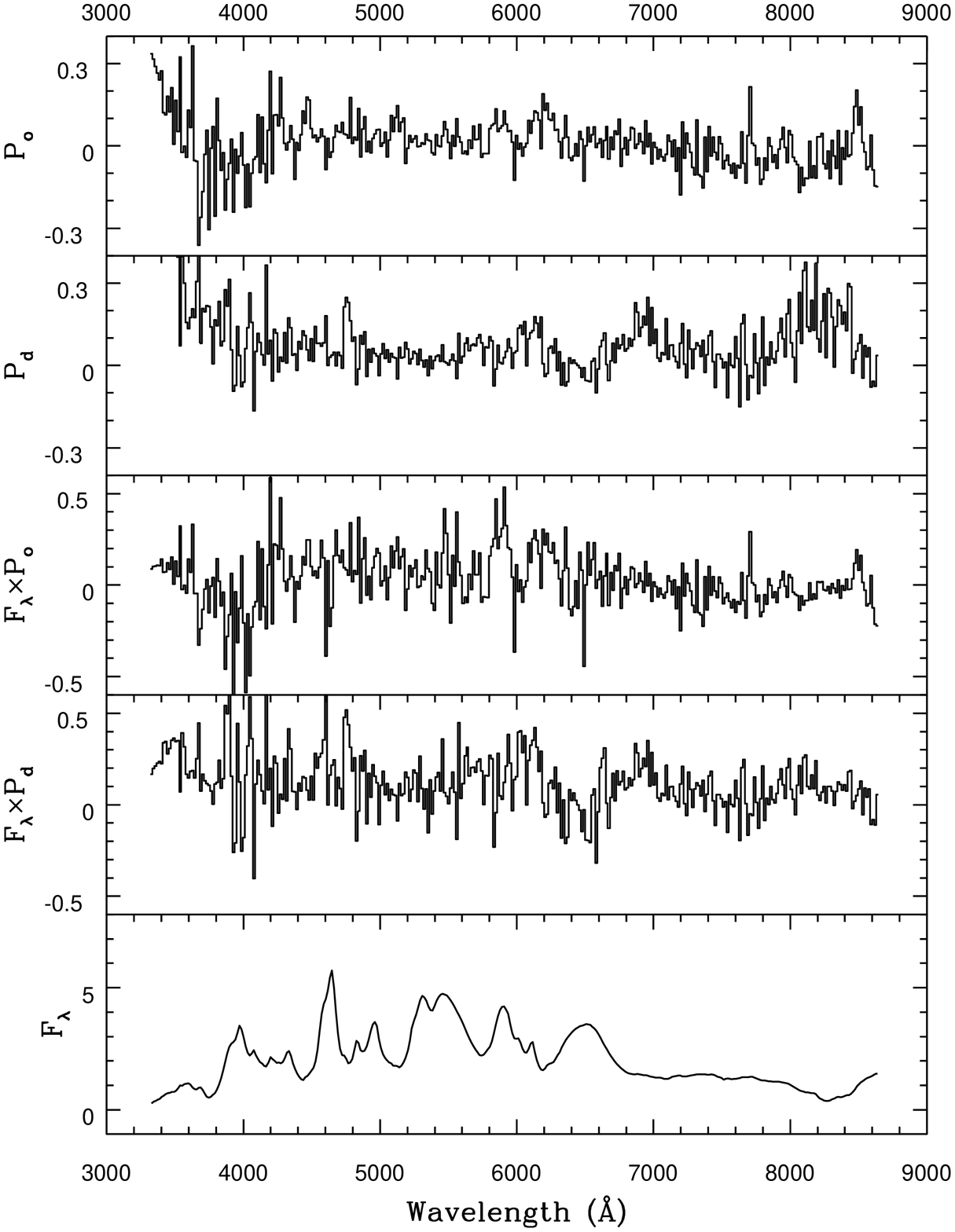}
\newpage
\plotone{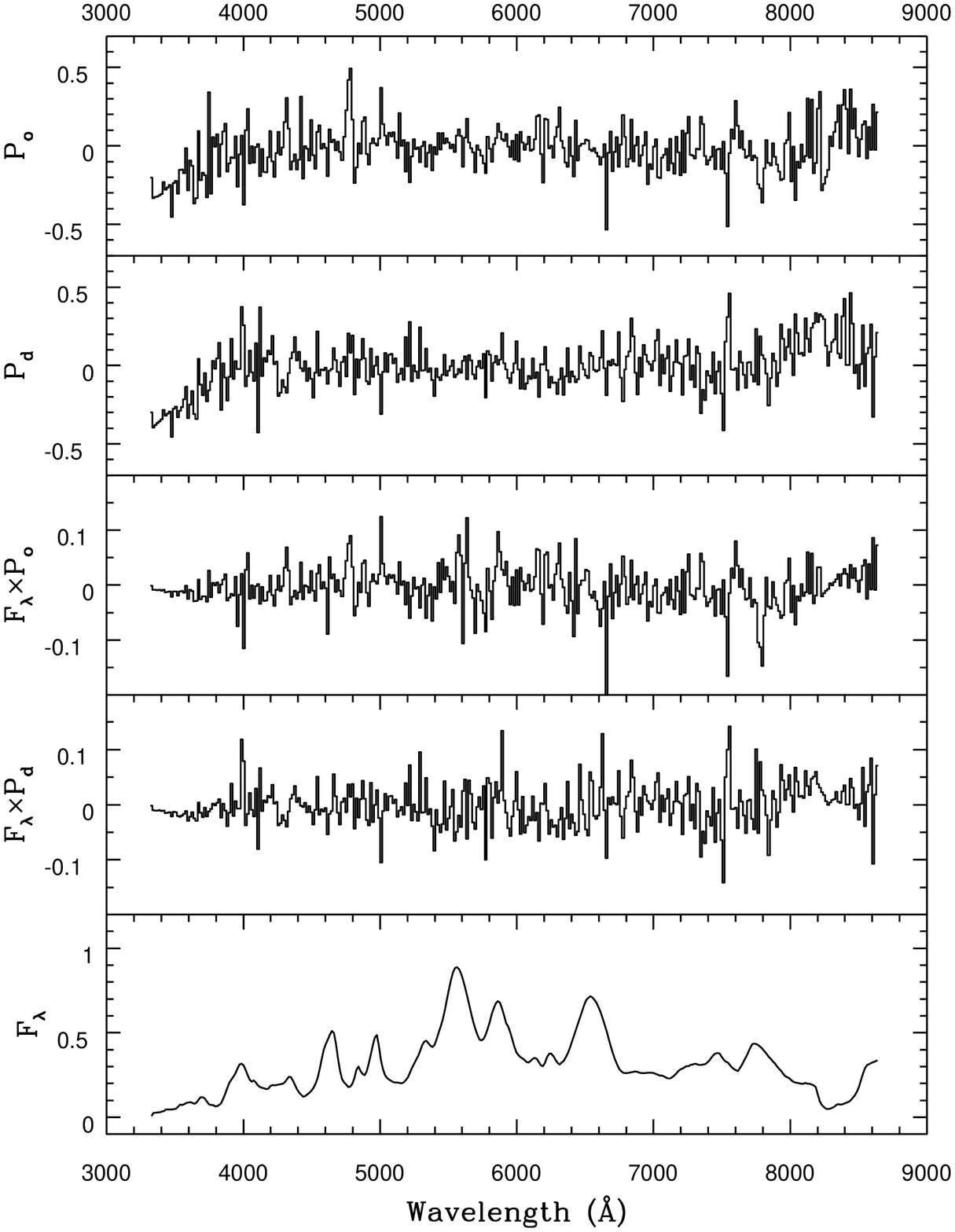}
\newpage
\plotone{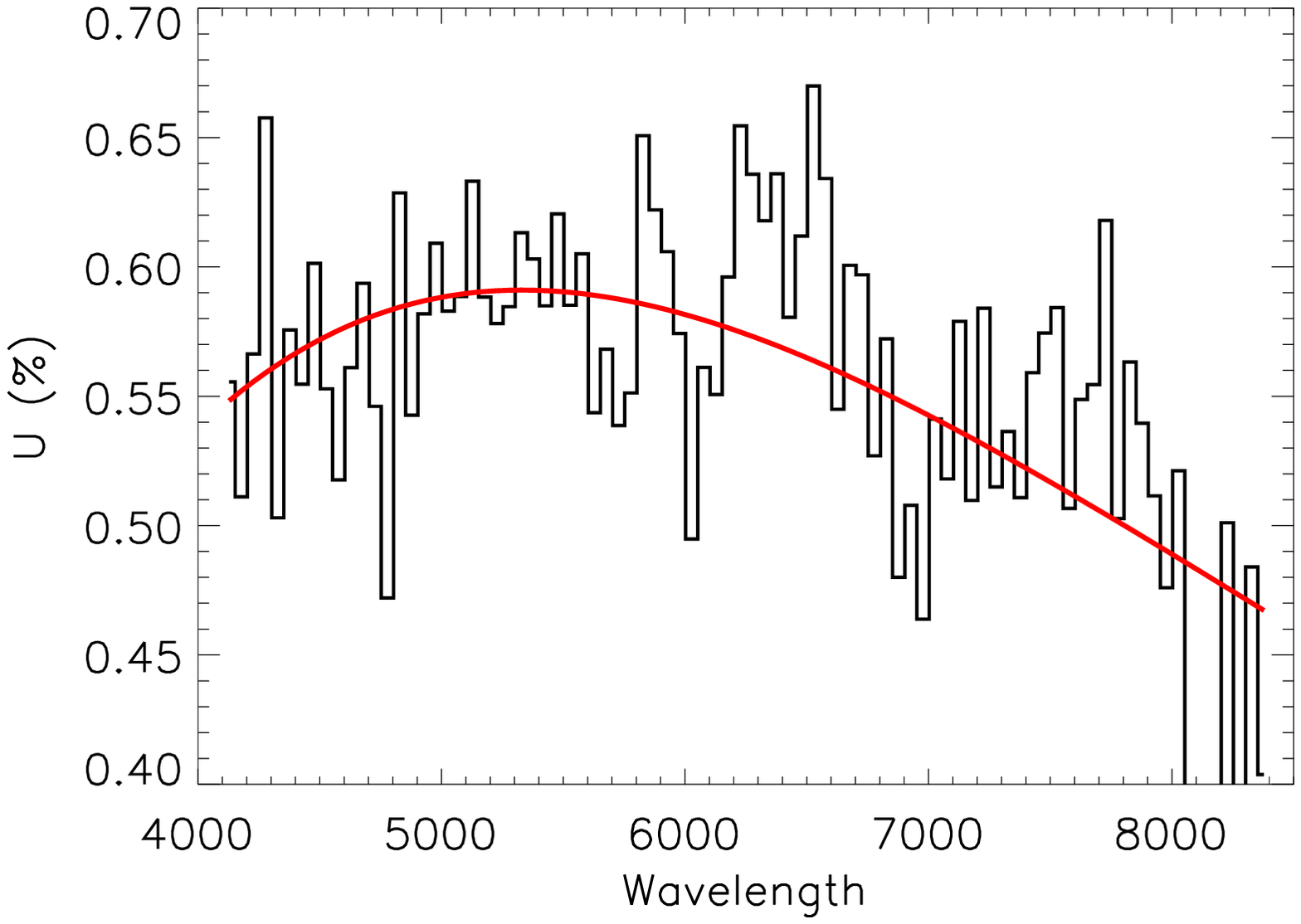}
\newpage
\epsscale{1.5}
\plotone{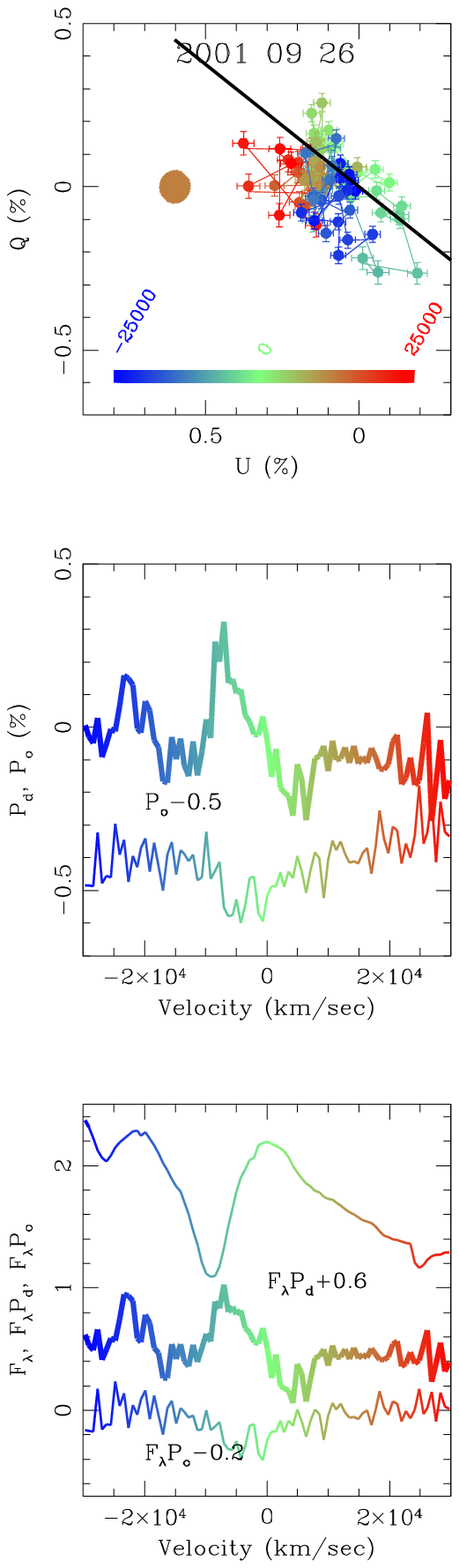}
\newpage
\plotone{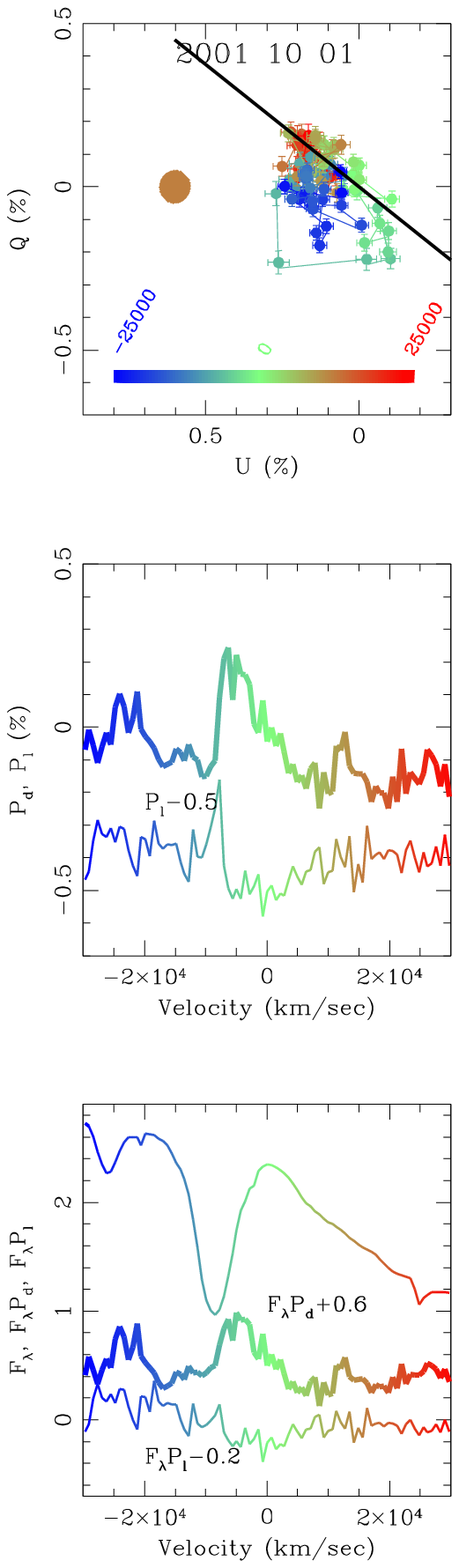}
\newpage
\plotone{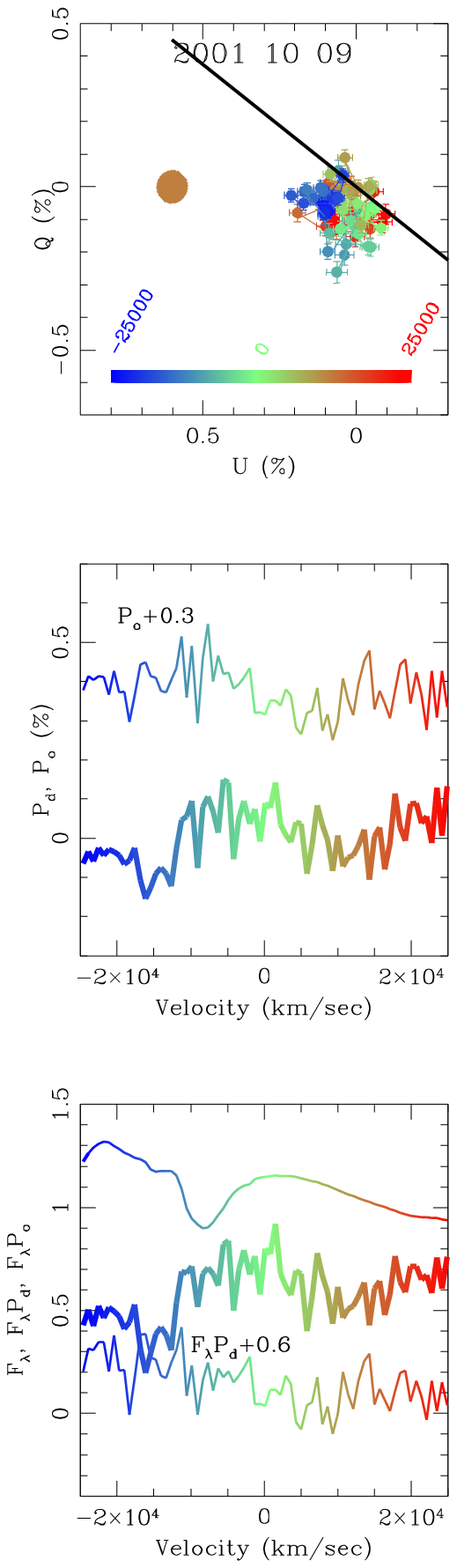}
\newpage
\plotone{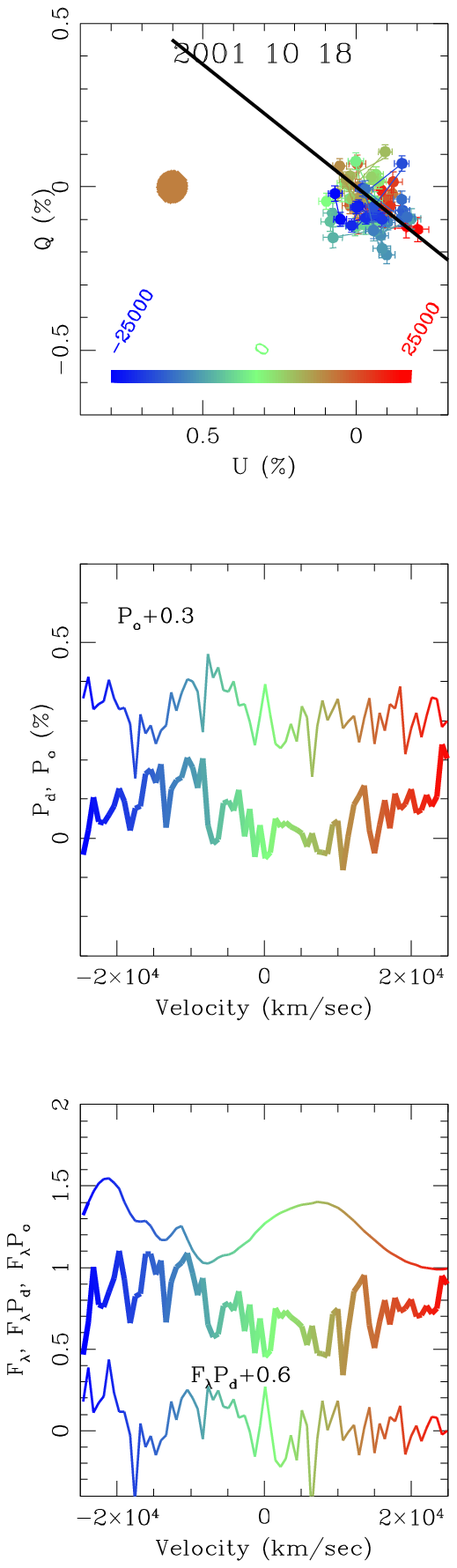}
\newpage
\plotone{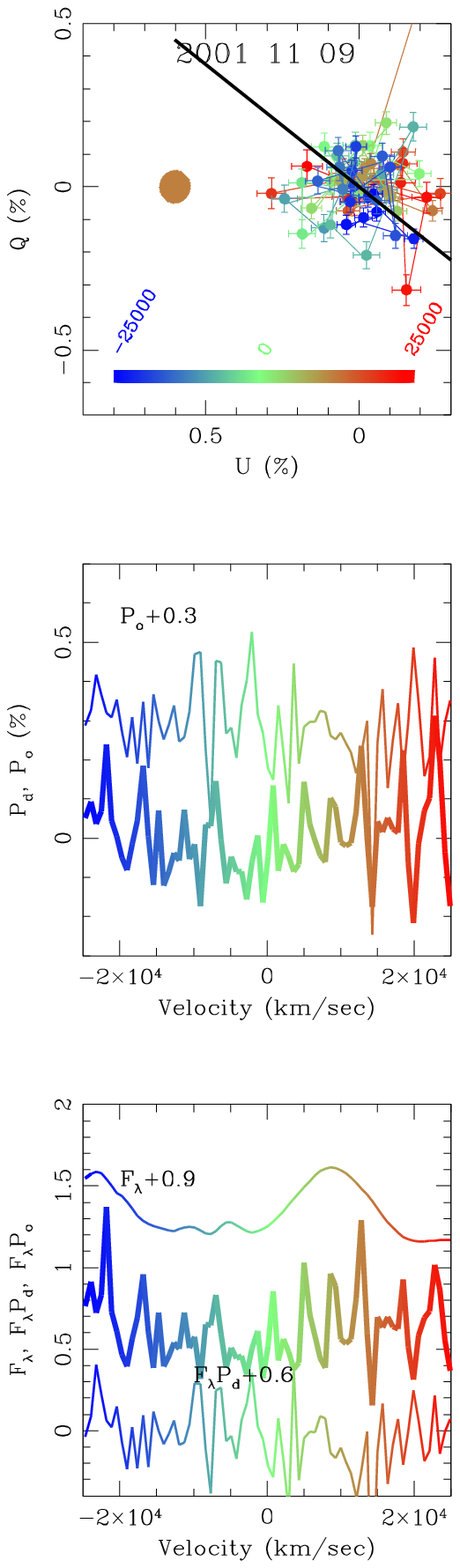}
\newpage
\plotone{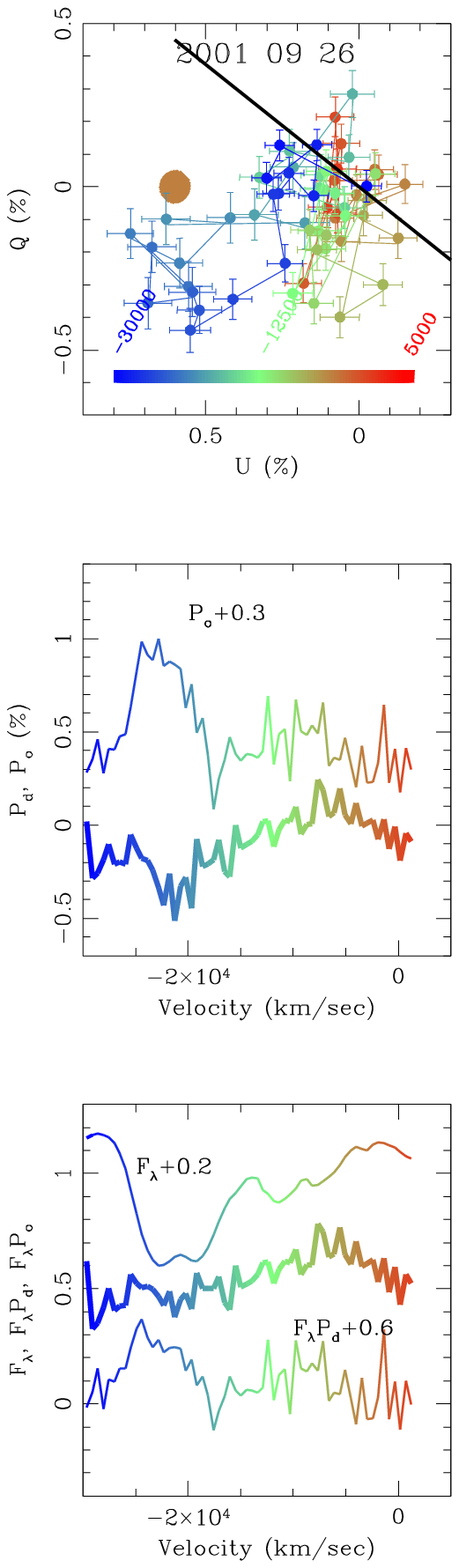}
\newpage
\plotone{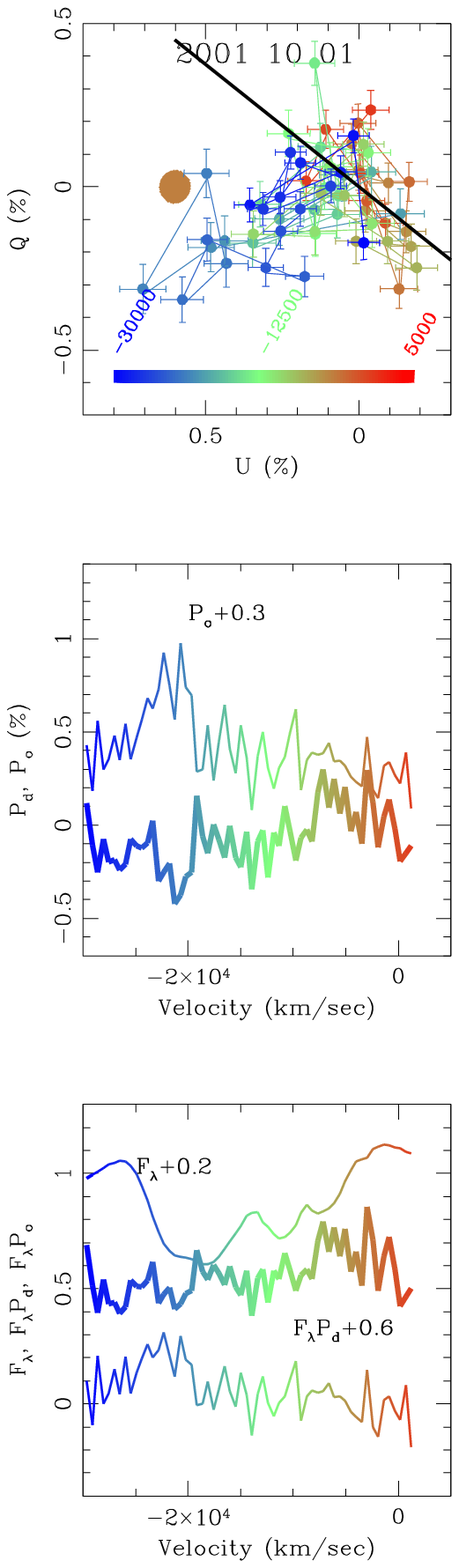}
\newpage
\plotone{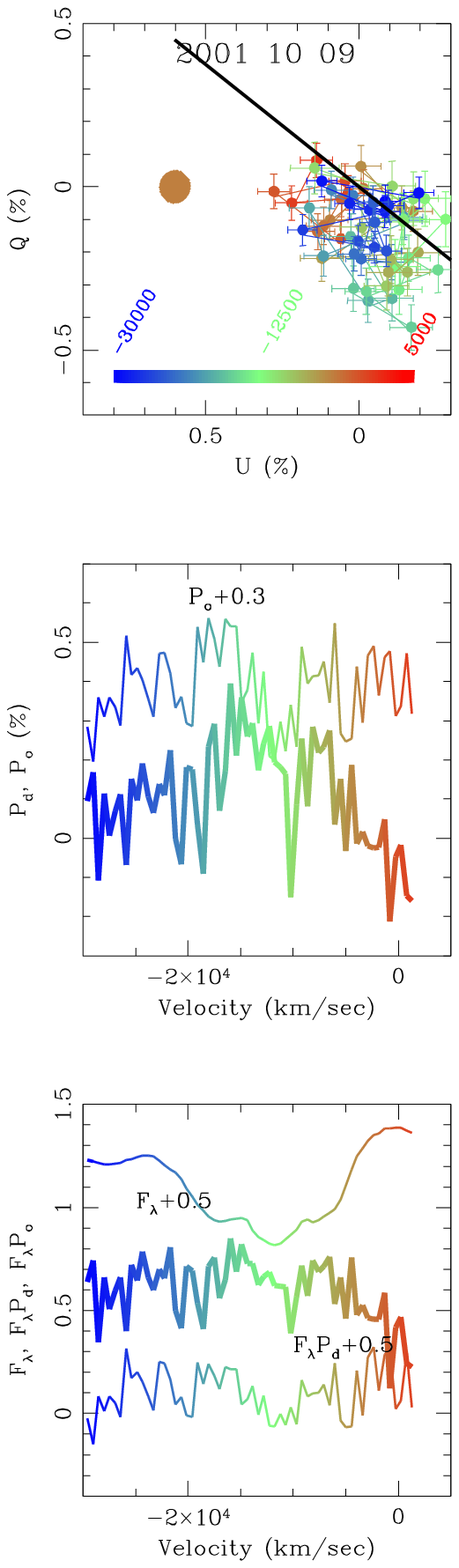}
\newpage
\plotone{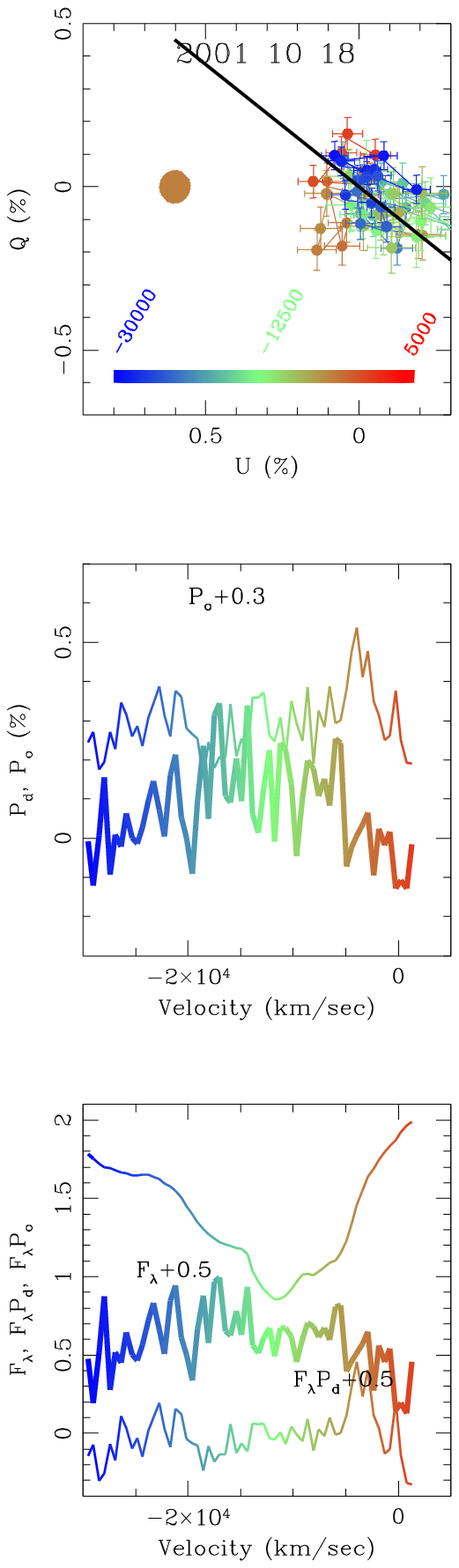}
\newpage
\plotone{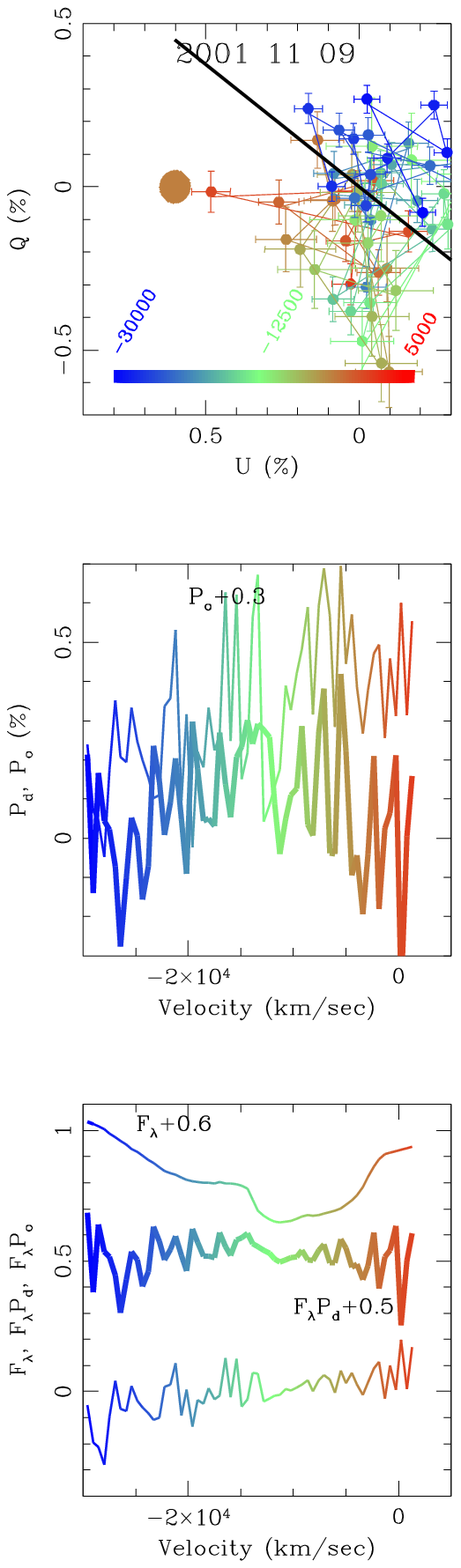}

\end{document}